\documentclass[12pt]{iopart}

\usepackage{graphicx}
\usepackage{color}
\usepackage[usenames,dvipsnames]{xcolor}
\usepackage{amssymb}
\usepackage[normalem]{ulem}
\usepackage{bm}

\makeatletter

\newcommand{\ie}{\textit{i.e.}}
\newcommand{\eg}{\textit{e.g.}}

\newtheorem{theo}{Theorem}

\newcommand{\sket}[1]{{\ensuremath{\lvert#1\rangle}}}
\newcommand{\lket}[1]{{\ensuremath{\left\lvert#1\right\rangle}}}
\newcommand{\ket}[1]{\if@display\lket{#1}\else\sket{#1}\fi}
\newcommand{\sbra}[1]{{\ensuremath{\langle#1\rvert}}}
\newcommand{\lbra}[1]{{\ensuremath{\left\langle#1\right\rvert}}}
\newcommand{\bra}[1]{\if@display\lbra{#1}\else\sbra{#1}\fi}
\newcommand{\sbraket}[2]{{\ensuremath{\langle#1\rvert#2\rangle}}}
\newcommand{\lbraket}[2]{{\ensuremath{\left\langle#1\!\left\rvert\vphantom{#1}#2\right.\!\right\rangle}}}
\newcommand{\braket}[2]{\if@display\lbraket{#1}{#2}\else\sbraket{#1}{#2}\fi}

\newcommand{\sketbra}[2]{{\ensuremath{\lvert #1\rangle\!\langle #2\rvert}}}
\newcommand{\lketbra}[2]{{\ensuremath{\left\lvert #1\right\rangle\!\!\left\langle #2\right\rvert}}}
\newcommand{\ketbra}[2]{\if@display\lketbra{#1}{#2}\else\sketbra{#1}{#2}\fi}
\makeatother
\begin{document}

\title[]{Practical aspects of measurement-device-independent quantum key distribution}

\author{Feihu Xu$^{1}$, Marcos Curty$^{2}$, Bing Qi$^{1,3}$ and Hoi-Kwong Lo$^{1}$}

\address{$^{1}$ Centre for Quantum Information and Quantum Control, Department of Physics and Department of Electrical \& Computer Engineering, University of Toronto, Toronto,  Ontario, M5S 3G4, Canada}
\address{$^{2}$ Escuela de Ingenier\'{\i}a de Telecomunicaci\'{o}n, Department of Signal Theory and Communications, University of Vigo, Vigo, Pontevedra, 36310, Spain}
\address{$^{3}$ Present address: Quantum Information Science Group, Computational Sciences and Engineering Division, Oak Ridge National Laboratory, Oak Ridge, Tennessee, 37831-6418, USA}

\ead{\mailto{feihu.xu@utoronto.ca}}

\begin{abstract}
A novel protocol, measurement-device-independent quantum key distribution (MDI-QKD), removes all attacks from the detection system, the most vulnerable part in QKD implementations. In this paper, we present an analysis for practical aspects of MDI-QKD. To evaluate its performance, we study various error sources by developing a general system model. We find that MDI-QKD is highly practical and thus can be easily implemented with standard optical devices. Moreover, we present a simple analytical method with only two (general) decoy states for the finite decoy-state analysis. This method can be used directly by experimentalists to demonstrate MDI-QKD. By combining the system model with the finite decoy-state method, we present a general framework for the optimal choice of the intensities of the signal and decoy states. Furthermore, we consider a common situation, namely \emph{asymmetric} MDI-QKD, in which the two quantum channels have different transmittances. We investigate its properties and discuss how to optimize its performance. Our work is of interest not only to experiments demonstrating MDI-QKD but also to other non-QKD experiments involving quantum interference.
\end{abstract}


\maketitle
\newpage
%%%%%%%%%%%%%%%%%%%%%%%%%%%%%%%%%%%%%%%%%%%%%%%%%%%%%%%%%%%%%%%%%%%%%%%%%%%%%%%%%%%%%%%

%%%%%%%%%%%%%%%%%%%%%%%%%%%%%%%%%%%%%%%%%%%%%%%%%%%%%%%%%%%%%%%%%%%%%%%%%%%%%%%%%%%%%%%
% Introduction
\section{Introduction}
Quantum key distribution (QKD)~\cite{bennett1984quantum,ekert1991quantum, QC:Gisin:2002} enables an unconditionally secure means of distributing secret keys between two spatially separated parties, Alice and Bob. The security of QKD has been rigorously proven based on the laws of quantum mechanics~\cite{security proof}. Nevertheless, owing to the imperfections in real-life implementations, a large gap between its theory and practice remains unfilled. In particular, an eavesdropper (Eve) may exploit these imperfections and launch specific attacks. This is commonly called quantum hacking. The first successful quantum hacking against a commercial QKD system was the time-shift attack~\cite{Yi:timeshift:2008} based on a proposal in~\cite{Qi:timeshift:2007}. More recently, the phase-remapping attack~\cite{phaseremapping} and the detector-control attack~\cite{detector:control} have been implemented against various practical QKD systems. Also, other attacks have appeared in the literature~\cite{other:attack}. These results suggest that quantum hacking is a major problem for the real-life security of QKD.

To close the gap between theory and practice, a natural attempt was to characterize the specific loophole and find a countermeasure. For instance, Yuan, Dynes and Shields proposed an efficient countermeasure against the detector-control attack~\cite{Yuan:counter}. Once an attack is known, the prevention is usually uncomplicated. However, unanticipated attacks are most dangerous, as it is impossible to fully characterize real devices and account for \emph{all} loopholes. Hence, researchers moved to the second approach -- (full) device-independent QKD~\cite{device:independent}. It requires no specification of the internal functionality of QKD devices and offers nearly perfect security. Its legitimate users (Alice and Bob) can be treated as a \emph{quasi} black box by assuming no memory attacks~\cite{barrett2013memory}. Nevertheless, device-independent QKD is not really practical because it requires near-unity detection efficiency and generates an extremely low key rate~\cite{device:independent2}. Therefore, to our knowledge, there has been no experimental paper on device-independent QKD.

\begin{figure}[!t]
\centering
\resizebox{8cm}{!}{\includegraphics{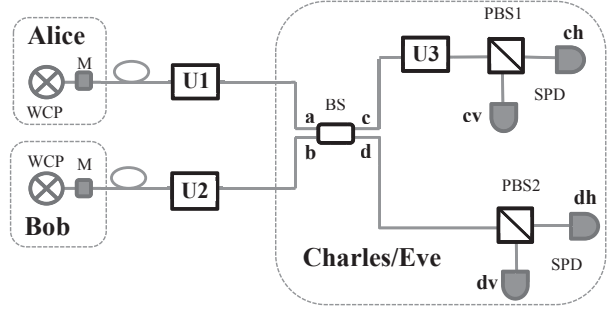}} \caption{\textbf{MDI-QKD system model.} WCP, weak coherent pulse; M, polarization and intensity modulators; BS, beam splitter; PBS, polarization beam splitter; SPD, single-photon detector. In MDI-QKD~\cite{Lo:MDIQKD}, each of Alice and Bob prepares BB84 states in combination with decoy states and sends them to an untrusted relay Charles (or Eve), who is supposed to perform a Bell state measurement. As an example, this figure considers a polarization-encoding scheme. Three unitary operators (U) are used to model the polarization misalignment (or rotation); PBS2 is defined as the fundamental measurement basis; $U_1$ ($U_2$) represents the polarization misalignment of Alice's (Bob's) channel transmission, while $U_3$ models the misalignment of the other measurement setting, PBS1. In the ideal case without any polarization misalignment, the rectilinear ($Z$) basis, used for key generation in Eq.~(\ref{Eqn:Key:formula}), refers to the basis of PBS1 and PBS2.} \label{Fig:model}
\end{figure}

Fortunately, Lo, Curty and Qi have recently proposed an innovative scheme -- measurement-device-independent QKD (MDI-QKD)~\cite{Lo:MDIQKD} -- that removes all detector side-channel attacks, the most important security loophole in conventional QKD implementations~\cite{Yi:timeshift:2008,Qi:timeshift:2007,detector:control, other:attack}. As an example of a MDI-QKD scheme (see Fig.~\ref{Fig:model}), each of Alice and Bob locally prepares phase-randomized signals (this phase randomization process can be realized using a quantum random number generator such as~\cite{xu2012ultrafast}) in the BB84 polarization states~\cite{bennett1984quantum} and sends them to an \emph{untrusted} quantum relay, Charles (or Eve). Charles is supposed to perform a Bell state measurement (BSM) and broadcast the measurement result. Since the measurement setting is only used to post-select entanglement (in an equivalent virtual protocol~\cite{Lo:MDIQKD}) between Alice and Bob, it can be treated as a \emph{true} black box. Hence, MDI-QKD is inherently immune to all attacks in the detection system. This is a major achievement as MDI-QKD allows legitimate users to not only perform secure quantum communications with untrusted relays~\footnote{This also implies the feasibility of ``Pentagon Using China Satellite for U.S.-Africa Command''. See http://www.bloomberg.com/news/2013-04-29/pentagon-using-china-satellite-for-u-s-africa-command.html.} but also out-source the manufacturing of detectors to untrusted manufactures.

Conceptually, the key insight of MDI-QKD is {\it time reversal}. This is in the same spirit as one-way quantum computation~\cite{oneway}. More precisely, MDI-QKD
built on the idea of a time-reversed EPR protocol for QKD~\cite{timereverse}. By combining the decoy-state method~\cite{decoyQKD} with the time-reversed EPR protocol, MDI-QKD gives both good performance and good security.

MDI-QKD is highly practical and can be implemented with standard optical components. The source can be a non-perfect single-photon source (together with the decoy-state method), such as an attenuated laser diode emitting weak coherent pulses (WCPs), and the measurement setting can be a simple BSM realized by linear optics. Hence, MDI-QKD has attracted intensive interest in the QKD community. A number of follow-up theoretical works have already been reported in~\cite{tamaki2012phase, PhysRevA.86.062319, ma2012statistical, wang2013three, song2012finite,Tittel:2012:MDI:theory, xu2013long}. Meanwhile, experimental attempts on MDI-QKD have also been made by several groups~\cite{Tittel:2012:MDI:exp, Liu:2012:MDI:exp, da2012proof, zhiyuan:experiment:2013}. Nonetheless, before it can be applied in real life, it is important to address a number of practical issues. These include:

\begin{figure}[!t]
\centering
\resizebox{6cm}{!}{\includegraphics{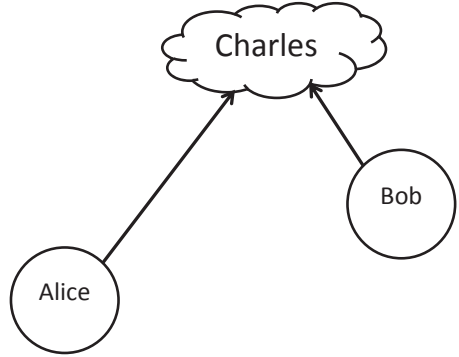}} \caption{\textbf{Asymmetric MDI-QKD.} The two channels connecting Alice to Charles and Bob to Charles have \emph{different} transmittances. In real-life MDI-QKD, asymmetry appeared naturally in a recent proof-of-concept experiment~\cite{Tittel:2012:MDI:exp}. } \label{Fig:general}
\end{figure}

\begin{enumerate}
  \item Modelling the errors: an implementation of MDI-QKD may involve various error sources such as the mode mismatch resulting in a non-perfect Hong-Ou-Mandel (HOM) interference~\cite{HOM:1987}. Thus, the first question is: how will these errors affect the performance of MDI-QKD~\cite{QBERorigin}? Or, what is the physical origin of the quantum bit error rate (QBER) in a practical implementation?

  \item Finite decoy-state protocol and finite-key analysis: as mentioned before, owing to the lack of true single-photon sources~\cite{grangier2004focus}, QKD implementations typically use laser diodes emitting WCPs~\cite{QC:Gisin:2002} and single-photon contributions are estimated by the decoy-state protocol~\cite{decoyQKD}. In addition, a real QKD experiment is completed in finite time, which means that the length of the output keys is finite. Thus, the estimation of relevant parameters suffers from statistical fluctuations. This is called the finite-key effect~\cite{finitekey}. Hence, the second question is: how can one design a practical finite decoy-state protocol and perform a finite-key analysis in MDI-QKD?

  \item Choice of intensities: an experimental implementation needs to know the optimal intensities for the signal and decoy states in order to optimize the system performance. Previously, \cite{ma2012statistical} and~\cite{wang2013three} have independently discussed the finite decoy-state protocol. However, the high computational cost of the numerical approach proposed in~\cite{ma2012statistical}, together with the lack of a rigorous discussion of the finite-key effect in both~\cite{ma2012statistical} and~\cite{wang2013three}, makes the optimization of parameters difficult. Thus, the third question is: how can one obtain these optimal intensities?

  \item Asymmetric MDI-QKD: as shown in Fig.~\ref{Fig:general}, in real life, it is quite common that the two channels connecting Alice and Bob to Charles have different transmittances. We call this situation \emph{asymmetric} MDI-QKD. Importantly, this asymmetric scenario appeared naturally in a recent proof-of-concept experiment~\cite{Tittel:2012:MDI:exp}, where a tailored length of fiber was intentionally added in the short arm to balance the two channel transmittances. Since an additional loss is introduced into the system, it is unclear whether this solution is optimal. Hence, the final question is: how can one optimize the performance of this asymmetric case?
\end{enumerate}

The second question has already been discussed in~\cite{ma2012statistical,wang2013three,song2012finite} and solved in ~\cite{marcos:finite:2013}. In this paper, we offer additional discussions on this point and answer the other questions. Our contributions are summarized below.

\begin{enumerate}
  \item To better understand the physical origin of the QBER, we propose generic models for various error sources. In particular, we investigate two important error sources -- polarization misalignment and mode mismatch. We find that in a polarization-encoding MDI-QKD system~\cite{Lo:MDIQKD, da2012proof, zhiyuan:experiment:2013}, polarization misalignment is the major source contributing to the QBER and mode mismatch (in the time or frequency domain), however, does not appear to be a major problem. These results are shown in Fig.~\ref{misalignment:asymptotic} and Fig.~\ref{timejitter:simulation}. Moreover, we provide a mathematical model to simulate a MDI-QKD system. This model is a useful tool for analyzing experimental results and performing the optimization of parameters. Although this model is proposed to study MDI-QKD, it is also useful for other non-QKD experiments involving quantum interference, such as entanglement swapping~\cite{sangouard2011quantum} and linear optics quantum computing~\cite{kok2007linear}. This result is shown in~\ref{App:QandE}.

  \item A previous method to analyze MDI-QKD with a finite number of decoy states assumes that Alice and Bob can prepare a vacuum state~\cite{wang2013three}. Here, however, we present an analytical approach with two \emph{general} decoy states, \ie, without the assumption of vacuum. This is particularly important for the practical implementations, as it is usually difficult to create a vacuum state in decoy-state QKD experiments~\cite{decoyexp,rosenberg2009practical}. The different intensities are usually generated with an intensity modulator, which has a finite extinction ratio (\eg, around 30 dB). Additionally, we also simulate the expected key rates numerically and thus present an optimized method with two decoy states. Ignoring for the moment the finite-key effect, experimentalists can directly use this method to obtain a rough estimation of the system performance. Table~\ref{Tab:finite:equations} contains the main results for this point.

  \item By combining the system model, the finite decoy-state protocol, and the finite-key analysis of~\cite{marcos:finite:2013}, we offer a general framework to determine the optimal intensities of the signal and decoy states. Notice that this framework has already been adopted and verified in the experimental demonstration reported in~\cite{zhiyuan:experiment:2013}. These results are shown in Fig.~\ref{Fig:flowchart} and Fig.~\ref{Fig:Optimal:intensity}.

  \item Finally, we model and evaluate the performance of an \emph{asymmetric} MDI-QKD system. This allows us to study its properties and determine the experimental configuration that maximizes its secret key rate. These results are shown in Table~\ref{Tab:Opt:Parameters} and Fig.~\ref{Fig:advantage:key}.
\end{enumerate}

%The rest of the paper is organized as follows. In Section~\ref{Sec:Preliminary}, we discuss the secure key rate and introduce the relevant notations. In Section~\ref{Sec:errors}, we present mathematical models to study various error sources. In Section~\ref{Sec:finitedecoy}, we analyze the finite decoy-state method with two general decoy states. We discuss the framework for finding the optimal intensities of signal and decoy states in Section~\ref{Sec:intensitychoice}. In Section~\ref{Sec:asymmetric}, we present the results for the asymmetric MDI-QKD. In Section~\ref{Sec:conclusion}, we provide some discussions about the assumptions in MDI-QKD and conclude this paper.
%%%%%%%%%%%%%%%%%%%%%%%%%%%%%%%%%%%%%%%%%%%%%%%%%%%%%%%%%%%%%%%%%%%%%%%%%%%%%
\section{Preliminary} \label{Sec:Preliminary}
The secure key rate of MDI-QKD in the asymptotic case (\ie, assuming an infinite number of decoy states and signals) is given by~\cite{Lo:MDIQKD}
\begin{equation} \label{Eqn:Key:formula}
    R\geq P^{1,1}_{Z}Y^{1,1}_{Z}[1-H_{2}(e_{X}^{1,1})]-Q_{Z}f_{e}(E_{Z})H_{2}(E_{Z}),
\end{equation}
where $Y_{Z}^{1,1}$ and $e_{X}^{1,1}$ are, respectively, the yield (\ie, the probability that Charles declares a successful event) in the rectilinear ($Z$) basis and the error rate in the diagonal ($X$) basis given that both Alice and Bob send single-photon states ($P^{1,1}_{Z}$ denotes this probability in the $Z$ basis); $H_{2}$ is the binary entropy function given by $H_2(x)$=$-x\log_2(x)-(1-x)\log_2(1-x)$; $Q_{Z}$ and $E_{Z}$ denote, respectively, the gain and QBER in the $Z$ basis and $f_{e}\geq 1$ is the error correction inefficiency function. Here we use the $Z$ basis for key generation and the $X$ basis for testing only~\cite{lo2005efficient}. In practice, $Q_{Z}$ and $E_{Z}$ are directly measured in the experiment, while $Y_{Z}^{1,1}$ and $e_{X}^{1,1}$ can be estimated using the finite decoy-state method.

Next, we introduce some additional notations. We consider one signal state and two weak decoy states for the finite decoy-state protocol. The parameter $\mu$ is the intensity (\ie, the mean photon number per optical pulse) of the signal state~\footnote{We assume that the coherent state is phase-randomized. Thus, its photon number follows a Poisson distribution of mean $\mu$.}. $\nu$ and $\omega$ are the intensities of the two decoy states, which satisfy $\mu>\nu>\omega\geq0$. The sets \{$\mu_a$,$\nu_a$,$\omega_a$\} and \{$\mu_b$,$\nu_b$,$\omega_b$\} contain respectively Alice's and Bob's intensities. The sets \{$\mu^{opt}_a$,$\nu^{opt}_a$,$\omega^{opt}_a$\} and \{$\mu^{opt}_b$,$\nu^{opt}_b$,$\omega^{opt}_b$\} denote the optimal intensities that maximize the key rate. $L_{ac}$ and $t_{a}$ ($L_{bc}$ and $t_{b}$) denote the channel distance and transmittance from Alice (Bob) to Charles. In the case of a fiber-based system, $t_{a}$=$10^{-\alpha L_{ac}/10}$ with $\alpha$ denoting the channel loss coefficient ($\alpha$$\approx$0.2 dB/km for a standard telecom fiber). $\eta_{d}$ is the detector efficiency and $Y_0$ is the background rate that includes detector dark counts and other background contributions. The parameters $e_{d}$, $e_{t}$, and $e_{m}$ denote, respectively, the errors associated with the polarization misalignment, the time-jitter and the total mode mismatch (see definitions below).

\section{Practical error sources} \label{Sec:errors}
In this section, we consider the original MDI-QKD setting~\cite{Lo:MDIQKD}, \ie, the symmetric case with $t_a$=$t_b$. The asymmetric case will be discussed in Section~\ref{Sec:asymmetric}. To model the practical error sources, we focus on the fiber-based polarization-encoding MDI-QKD system proposed in~\cite{Lo:MDIQKD} and demonstrated in~\cite{da2012proof, zhiyuan:experiment:2013}. Notice, however, that with some modifications, our analysis can also be applied to other implementations such as free-space transmission, the phase-encoding scheme and the time-bin-encoding scheme. See also~\cite{PhysRevA.86.062319} and~\cite{Tittel:2012:MDI:theory} respectively for models of phase-encoding and time-bin-encoding schemes.

A comprehensive list of practical error sources is as follows~\footnote{This list does not consider the state-preparation error~\cite{tamaki2012phase,wang2013three}, because a strict discussion about this problem is related to the security proof of MDI-QKD, which will be considered in future publications.}.
\begin{enumerate}
  \item Polarization misalignment (or rotation).
  \item Mode mismatch including time-jitter, spectral mismatch and pulse-shape mismatch.
  \item Fluctuations of the intensities (modulated by Alice and Bob) at the source.
  \item Background rate.
  \item Asymmetry of the beam splitter.
\end{enumerate}

Here, we primarily analyze the first two error sources, \ie, polarization misalignment and mode mismatch. The other error sources present minor contributions to the QBER in practice, and are discussed in~\ref{App:othererrors}.

\subsection{Polarization misalignment} \label{Sec:polarization}
Polarization misalignment (or rotation) is one of the most significant factors contributing to the QBER in not only the polarization-encoding BB84 system~\cite{bennett1984quantum} but also the polarization-encoding MDI-QKD system. Since MDI-QKD requires two transmitting channels and one BSM (instead of one channel and a simple measurement as in the BB84 protocol), it is cumbersome to model its polarization misalignment. Here, we solve this problem by proposing a simple model in Fig.~\ref{Fig:model}. One of the polarization beam splitters (PBS2 in Fig.~\ref{Fig:model}) is defined as the fundamental measurement basis~\footnote{Although we use PBS2 as the reference basis, the method is also applicable to other reference bases such as PBS1.}. Three unitary operators, \{$U_1$, $U_2$, $U_3$\}, are considered to model the polarization misalignment of each channel~\cite{kok2007linear}. The operator $U_1$ ($U_2$) represents the misalignment of Alice's (Bob's) channel transmission, while $U_3$ models the misalignment of the other measurement setting, PBS1.

For simplicity, we consider a simplified model with a 2-dimensional unitary matrix~\footnote{That is, if we denote the two incoming modes in the horizontal and vertical polarization by the creation operators $a_{h}^{\dagger}$ and $a_{v}^{\dagger}$, and the outgoing modes by $b_{h}^{\dagger}$ and $b_{v}^{\dagger}$, then the unitary operator yields an evolution of the form $b_{h}^{\dagger}=\cos\theta_{k}a_{h}^{\dagger}-\sin\theta_{k}a_{v}^{\dagger}$ and $b_{v}^{\dagger}=\sin\theta_{k}a_{h}^{\dagger}+\cos\theta_{k}a_{v}^{\dagger}$. This unitary matrix is a simple form rather than the general one (see Section I.A in~\cite{kok2007linear}). Nonetheless, we believe that the result for a more general unitary transformation will be similar to our simulation results.}
\begin{equation}\label{Unitary}
  U_{k}=\left(
  \begin{array}{cc}
    \cos\theta_{k} & -\sin\theta_{k} \\
    \sin\theta_{k} & \cos\theta_{k}
  \end{array}
\right),
\end{equation}
where $k$=1, 2, 3 and $\theta_{k}$ (polarization-rotation angle) is in the range of [$-\pi$,$\pi$]. For each value of $k$, we define the polarization misalignment error $e_{k}$=$\sin^{2}\theta_{k}$ and the total error $e_d$=$\sum_{k=1}^{3}e_k$. Note that $e_{d}$ is equivalent to the systematic QBER in a polarization-encoding BB84 system.

\begin{table}[hbt]
\centering
\begin{tabular}{c @{\hspace{0.5cm}} c @{\hspace{0.5cm}} c @{\hspace{0.5cm}} c @{\hspace{0.5cm}} c} \hline
$\eta_{d}$ & $e_{d}$ & $Y_{0}$ & $f_{e}$ & $e_{m}$ \\
\hline
14.5\% & 1.5\% & $6.02\times 10^{-6}$  & 1.16 & 2\% \\
\hline
\end{tabular}
\caption{List of practical parameters for all numerical simulations. These experimental parameters, including the detection efficiency $\eta_{d}$, the total misalignment error $e_{d}$ and the background rate $Y_{0}$, are from the 144km QKD experiment reported in~\cite{Ursin:144QKD}. Since two SPDs are used in~\cite{Ursin:144QKD}, the background rate of each SPD here is roughly half of the value there. We assume that the four SPDs in MDI-QKD (see Fig.~\ref{Fig:model}) have identical $\eta_{d}$ and $Y_{0}$. The parameter $e_{m}$ is the total mode mismatch that is quantified from the experimental values of~\cite{zhiyuan:experiment:2013}.} \label{Tab:exp:parameters}
\end{table}

\begin{figure}[!t]
\centering
\resizebox{7cm}{!}{\includegraphics{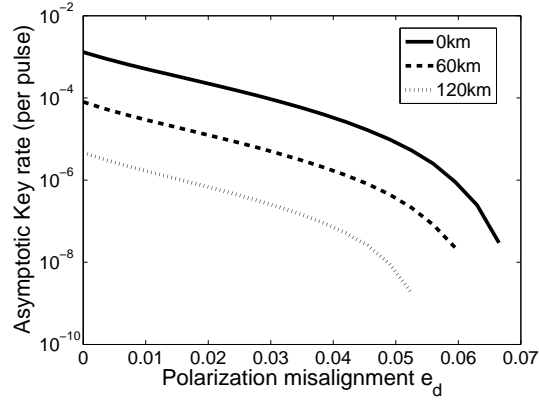}} \caption{\textbf{Polarization misalignment tolerance.} Following the model illustrated in Fig.~\ref{Fig:model}, we incorporate the polarization misalignment into the derivation of the asymptotic key rate given by Eq.~(\ref{Eqn:Key:formula}). We find that MDI-QKD is robust against practical errors due to polarization misalignment.} \label{misalignment:asymptotic}
\end{figure}

From the model of Fig.~\ref{Fig:model}, we can analyze the effect of polarization misalignment by evaluating the secure key rate given by Eq.~(\ref{Eqn:Key:formula}). See~\ref{App:QandE} for details. By using the practical parameters listed in Table~\ref{Tab:exp:parameters}, we perform a numerical simulation of the asymptotic key rates for different values of polarization misalignment, $e_d$. The result is shown in Fig.~\ref{misalignment:asymptotic}. In this simulation, we temporarily ignore mode mismatch (\ie, set $e_m$=0 in Table~\ref{Tab:exp:parameters}) and make two practical assumptions for the polarization misalignment: a) each polarization-rotation angle, $\theta_{k}$, follows a \emph{Gaussian} distribution with a standard deviation of $\theta_{k}^{std}=\arcsin(\sqrt{e_{k}})$; and b) the probability distribution of $e_k$ is selected as $e_1$=$e_2$=0.475$e_{d}$ and $e_3$=0.05$e_{d}$~\footnote{Two remarks for the distribution of the three unitary operators: a) We assume that the two channel transmissions, \ie, $U_1$ and $U_2$, introduce much larger polarization misalignments than the other measurement basis, $U_3$ (PBS1 in Fig.~\ref{Fig:model}), because PBS1 is located in Charles's local station and can be carefully aligned (in principle). Hence, we choose $e_1$=$e_2$=0.475$e_{d}$ and $e_3$=0.05$e_{d}$. b) Notice that the simulation result is more or less \emph{independent} of the distribution of $e_k$.}. Fig.~\ref{misalignment:asymptotic} shows that a polarization-encoding system can tolerate up to about 6.7\% polarization misalignment at 0 km, while at 120 km it can only tolerate up to 5\% misalignment. It also shows that MDI-QKD is moderately robust to errors due to polarization misalignment.

%%%%%%%%%%%%%%%%%%%%%%%%%%%%%%%%%%%%%%%%%%%%%%%%%%%%%%%%%%%%%%%%%%%%%%%%%%%%%%%%%%%%%%%%
\subsection{Mode mismatch} \label{timejitter:section}
\begin{figure}[!t]
\centering
\resizebox{8cm}{!}{\includegraphics{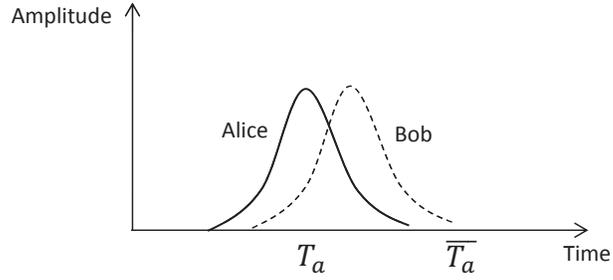}} \caption{\textbf{Model for mode mismatch in the time domain (time-jitter).} Alice's state is defined as the reference basis, while Bob's state is a superposition of Alice's fundamental mode $|T_{a}\rangle$ and the orthogonal mode $|\overline{T_{a}}\rangle$ (see Eq.~(\ref{Eqn:timejitter:model})).} \label{Fig:model:jitter}
\end{figure}

We primarily use the model of mode mismatch in time domain, called time-jitter~\footnote{Time-jitter is the variance in arrival times of Alice's and Bob's packets at Charles's station.}, to discuss our method. This model is shown in Fig.~\ref{Fig:model:jitter}. We describe Alice's and Bob's quantum states in the time domain as
\begin{eqnarray} \label{Eqn:timejitter:model}
 \nonumber % to remove numbering (before each equation)
  &Alice&: |\phi_{a}\rangle=|T_{a}\rangle \\
  &Bob&: |\phi_{b}\rangle=\alpha|T_{a}\rangle+\beta|\overline{T_{a}}\rangle,
\end{eqnarray}
where $|\overline{T_{a}}\rangle$ is the orthogonal time mode of $|T_{a}\rangle$, $\beta$=$\sqrt{e_{t}}$, $\alpha$=$\sqrt{1-e_{t}}$, and $e_{t}$ is defined as the time-jitter that represents the probability of Alice's state not overlapping with that of Bob~\footnote{In experiment, the value of $e_{t}$ can be quantified from the fidelity between the two pulses in time domain. This fidelity can be obtained by measuring the pulse width and the time-jitter value between the two pulses. From the experimental values of~\cite{zhiyuan:experiment:2013}, $e_t$ is below 1.5\%.}.

This model is a very general method that can be used to study the mode mismatch problem in other domains for a variety of quantum optics experiments involving quantum interference. For instance, a similar discussion can be applied to the spectral (wavelength) mismatch if we write Eq.~(\ref{Eqn:timejitter:model}) in the frequency domain. One can also refer to~\cite{rohde2007spectral} for a general discussion about the spectral mismatch. Considering Eq.~(\ref{Eqn:timejitter:model}) in the form of Alice's and Bob's pulse shapes, we can also analyze the pulse-shape mismatch. Here we define the \emph{total} mode mismatch in all domains as $e_m$.

Next, let us discuss how $e_m$ affects the key rate given by Eq.~(\ref{Eqn:Key:formula}). As illustrated in Fig.~\ref{Fig:model}, the overlapping modes between Alice's and Bob's pulses experience a HOM interference at the beam splitter (BS), while the non-overlapping modes transmit through the BS without interference. Assuming that $\eta_d\gg Y_{0}$ and ignoring the polarization misalignment for the moment, we find that the mode mismatch \emph{only} affects the gains and the error rates in the $X$ basis rather than those in the $Z$ basis~\cite{mismatchonkey}. Hence, in Eq.~(\ref{Eqn:Key:formula}), $e_m$ mainly affects $e_{X}^{1,1}$. In practice, $e_{X}^{1,1}$ can be estimated from the finite decoy-state protocol, \ie, from the gains ($Q_{X}$) and QBERs ($E_{X}$) in the $X$ basis. Similar to the analysis of the polarization misalignment in Section~\ref{Sec:polarization}, we can incorporate $e_m$ into the derivations of $Q_{X}$ and $E_{X}$ following the method of~\ref{App:QandE}~\cite{QxEx}.
\begin{figure}[!t]
\centering
\resizebox{7cm}{!}{\includegraphics{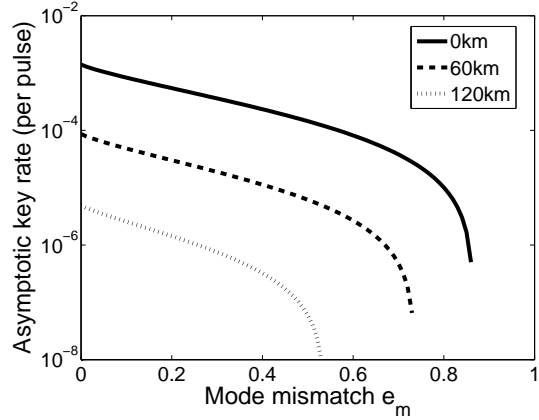}} \caption{\textbf{Mode mismatch tolerance.} In the asymptotic case, a polarization-encoding MDI-QKD system can tolerate up to 80\% mode mismatch at 0 km. Mode mismatch does not appear to be a major problem in a polarization-encoding implementation of MDI-QKD.} \label{timejitter:simulation}
\end{figure}

Using the parameters of Table~\ref{Tab:exp:parameters}, we simulate the asymptotic key rates for different values of $e_m$. The results are shown in Fig.~\ref{timejitter:simulation}. In this simulation, we temporarily ignore polarization misalignment (\ie, we set $e_d$=0) and only focus on mode mismatch. At 0 km, we find that the system can tolerate up to 80\% mode mismatch and at 120 km, the tolerable value is about 50\%. Hence, a polarization-encoding MDI-QKD system is \emph{less} sensitive to mode mismatch than to polarization misalignment~\cite{timebin}. Notice also that we have quantified the value of $e_m$ (see Table~\ref{Tab:exp:parameters}) by using the experimental parameters from~\cite{zhiyuan:experiment:2013} and find that $e_m$ is usually small in practice (\eg, below 5\%). Therefore, mode mismatch does not appear to be a major problem in a MDI-QKD implementation.

\section{Finite decoy-state protocol with two general decoy states} \label{Sec:finitedecoy}
In a MDI-QKD implementation, by performing the measurements for the different intensities used by Alice and Bob, we can obtain~\cite{Lo:MDIQKD}
\begin{equation} \label{Eqn:general:decoy1}
Q^{q_aq_b}_{Z}=\sum_{n,m=0}e^{-(q_a+q_b)}\frac{q^n_a}{n!}\frac{q^m_b}{m!}Y^{n,m}_{Z},
\end{equation}
\begin{equation} \label{Eqn:general:decoy2}
Q^{q_aq_b}_{X}E^{q_aq_b}_{X}=\sum_{n,m=0}e^{-(q_a+q_b)}\frac{q^n_a}{n!}\frac{q^m_b}{m!}Y^{n,m}_{X}e^{n,m}_{X},
\end{equation}
where $q_a$ ($q_b$) denotes Alice's (Bob's) intensity setting, $Q^{q_aq_b}_{Z}$ ($E^{q_a, q_b}_{X}$) denotes the gain (QBER) in the $Z$ ($X$) basis with the intensity pair \{$q_a$, $q_b$\}, and $Y^{n,m}_{Z}$ ($e^{n,m}_{X}$) denotes the yield (error rate) given that Alice and Bob send respectively an $n$-photon and $m$-photon pulse. Here the goal of the finite decoy-state protocol is to estimate $Y_{Z}^{1,1}$ and $e_{X}^{1,1}$ (used to generate a secure key) from the set of linear equations given by Eqs.~(\ref{Eqn:general:decoy1}) and~(\ref{Eqn:general:decoy2}) using different intensity settings~\cite{twodecoy}. More specifically, we estimate a lower bound for $Y_{Z}^{1,1}$ and an upper bound for $e_{X}^{1,1}$. We denote these two bounds respectively as $Y_{Z,L}^{1,1}$ and $e_{X,U}^{1,1}$.

\begin{table}[hbt]
\centering
\scalebox{1}{%
\begin{tabular}{c}
\hline
$Y_{Z,L}^{1,1}\geq\frac{1}{(\mu_a-\omega_a)(\mu_b-\omega_b)(\nu_a-\omega_a)(\nu_b-\omega_b)(\mu_a-\nu_a)} \times$ \\ $[(\mu_a^2-\omega_a^2)(\mu_b-\omega_b)(Q^{\nu_a\nu_b}_{Z}e^{(\nu_a+\nu_b)}+Q^{\omega_a\omega_b}_{Z}e^{(\omega_a+\omega_b)}-Q^{\nu_a\omega_b}_{Z}e^{(\nu_a+\omega_b)}-Q^{\omega_a\nu_b}_{Z}e^{(\omega_a+\nu_b)})-$ \\
$(\nu_a^2-\omega_a^2)(\nu_b-\omega_b)(Q^{\mu_a\mu_b}_{Z}e^{\mu_a+\mu_b}+Q^{\omega_a\omega_b}_{Z}e^{(\omega_a+\omega_b)}-Q^{\mu_a\omega_b}_{Z}e^{(\mu_a+\omega_b)}-Q^{\omega_a\mu_b}_{Z}e^{(\omega_a+\mu_b)}]$ \\
\hline
$e_{X,U}^{1,1}\geq\frac{1}{(\nu_a-\omega_a)(\nu_b-\omega_b)Y_{X,L}^{1,1}} \times$ \\
$[e^{(\nu_a+\nu_b)}Q^{\nu_a\nu_b}_{X}E^{\nu_a\nu_b}_{X}+e^{(\omega_a+\omega_b)}Q^{\omega_a\omega_b}_{X}E^{\omega_a\omega_b}_{X}-
e^{(\nu_a+\omega_b)}Q^{\nu_a\omega_b}_{X}E^{\nu_a\omega_b}_{X}-e^{(\omega_a+\nu_b)}Q^{\omega_a\nu_b}_{X}E^{\omega_a\nu_b}]$
\\
\hline
\end{tabular}}
\caption{\textbf{Analytical equations for the two decoy-state protocol.} See the main text for details. Here, for the estimation of $Y_{Z,L}^{1,1}$, we only consider one case and refer to Eq.~(\ref{Y11:2decoy:final2}) for the other case. Ignoring the finite-key effect, these results can be directly used by experimentalists to obtain an estimation of the expected system performance.} \label{Tab:finite:equations}
\end{table}

The general approach for the finite decoy-state protocol has been discussed in~\cite{marcos:finite:2013}. In this section, however, we present a much simpler analytical method with only two decoy states. The final results are summarized in Table~\ref{Tab:finite:equations}. They can be directly used by experimentalists (without knowing the details of~\cite{marcos:finite:2013}) to obtain a rough estimation of the expected system performance. Notice that our notations are different from~\cite{marcos:finite:2013} in that we primarily estimate the probabilities in the case of an infinite number of signals, while~\cite{marcos:finite:2013} focuses on the estimation of counts by incorporating the finite-key effect.

Now, let us start to discuss this two decoy-state protocol. As mentioned before, the intensities of the signal and decoy states satisfy $\mu>\nu>\omega$ and our protocol is applicable to either $\omega=0$ or $\omega\neq0$. The key method to estimate $Y_{Z,L}^{1,1}$ from Eq.~(\ref{Eqn:general:decoy1}) can be divided into two steps:
\begin{enumerate}
  \item Cancel out the terms $Y^{0m}_{Z}$ and $Y^{n0}_{Z}$ using Gaussian elimination.
  \item Cancel out either the term $Y^{12}_{Z}$ or $Y^{21}_{Z}$ depending on the intensity values selected in the first step.
\end{enumerate}

For the first step, we choose intensity pairs from \{$\mu_a$, $\omega_a$, $\mu_b$, $\omega_b$\} and \{$\nu_a$, $\omega_a$, $\nu_b$, $\omega_b$\}~\cite{generalintensity1}, and generate two quantities $Q_{Z}^{M1}$ and $Q_{Z}^{M2}$ given by
\begin{eqnarray} \label{Y11:2decoy:QrectM} \nonumber
Q_{Z}^{M1}=Q^{\nu_a\nu_b}_{Z}e^{(\nu_a+\nu_b)}+ Q^{\omega_a\omega_b}_{Z}e^{(\omega_a+\omega_b)}-Q^{\nu_a\omega_b}_{Z}e^{(\nu_a+\omega_b)}-Q^{\omega_a\nu_b}_{Z}e^{(\omega_a+\nu_b)}, \\ \nonumber
Q_{Z}^{M2}=Q^{\mu_a\mu_b}_{Z}e^{(\mu_a+\mu_b)}+Q^{\omega_a\omega_b}_{Z}e^{(\omega_a+\omega_b)}-Q^{\mu_a\omega_b}_{Z}e^{(\mu_a+\omega_b)}-Q^{\omega_a\mu_b}_{Z}e^{(\omega_a+\mu_b)}.
\end{eqnarray}

To cancel out $Y^{12}_{Z}$ or $Y^{21}_{Z}$, we consider two cases.
\begin{description}
  \item[Case 1. ($\frac{\mu_a+\omega_a}{\nu_a+\omega_a}\leq \frac{\mu_b+\omega_b}{\nu_b+\omega_b}$)]  we use $(\mu_b^2-\omega_b^2)(\mu_a-\omega_a)\times Q_{Z}^{M1}$ minus $(\nu_b^2-\omega_b^2)(\nu_a-\omega_a)\times Q_{Z}^{M2}$ to cancel out $Y^{12}_{Z}$. Thus, $Y_{Z,L}^{1,1}$ is given by
\begin{equation} \label{Y11:2decoy:final1}
\frac{(\mu_a^2-\omega_a^2)(\mu_b-\omega_b)Q_{Z}^{M1}-(\nu_a^2-\omega_a^2)(\nu_b-\omega_b)Q_{Z}^{M2}}{(\mu_a-\omega_a)(\mu_b-\omega_b)(\nu_a-\omega_a)(\nu_b-\omega_b)(\mu_a-\nu_a)}.
\end{equation}
  \item[Case 2. ($\frac{\mu_a+\omega_a}{\nu_a+\omega_a}>\frac{\mu_b+\omega_b}{\nu_b+\omega_b}$)] we cancel out $Y^{21}_{Z}$ using the same method as in case 1 and derive $Y_{Z,L}^{1,1}$ as
\begin{equation} \label{Y11:2decoy:final2}
\frac{(\mu_b^2-\omega_b^2)(\mu_a-\omega_a)Q_{Z}^{M1}-(\nu_b^2-\omega_b^2)(\nu_a-\omega_a)Q_{Z}^{M2}}{(\mu_a-\omega_a)(\mu_b-\omega_b)(\nu_a-\omega_a)(\nu_b-\omega_b)(\mu_b-\nu_b)}.
\end{equation}
\end{description}

Similarly, the strategy to estimate $e_{X,U}^{1,1}$ from Eq.~(\ref{Eqn:general:decoy2}) requires to cancel out $Y^{0,m}_{X}e^{0,m}_{X}$ and $Y^{n,0}_{X}e^{n,0}_{X}$. Thus, we choose intensity pairs from \{$\nu_a$, $\omega_a$, $\nu_b$, $\omega_b$\}~\cite{generalintensity2} and derive $e_{X,U}^{1,1}$ as
\begin{eqnarray} \label{e11:2decoy:final} \nonumber
\frac{1}{(\nu_a-\omega_a)(\nu_b-\omega_b)Y_{X,L}^{1,1}}&[e^{(\nu_a+\nu_b)}Q^{\nu_a\nu_b}_{X}E^{\nu_a\nu_b}_{X}+e^{(\omega_a+\omega_b)}Q^{\omega_a\omega_b}_{X}E^{\omega_a\omega_b}_{X} \\ \nonumber
&-e^{(\nu_a+\omega_b)}Q^{\nu_a\omega_b}_{X}E^{\nu_a\omega_b}_{X}-e^{(\omega_a+\nu_b)}Q^{\omega_a\nu_b}_{X}E^{\omega_a\nu_b}].
\end{eqnarray}
where $Y_{X,L}^{1,1}$ can be estimated using a similar method to that for $Y_{Z,L}^{1,1}$~\cite{Yx11}. The final equations are summarized in Table~\ref{Tab:finite:equations}.

\section{Optimal choice of intensities} \label{Sec:intensitychoice}
\begin{figure}[!t]
\centering
\resizebox{7.5cm}{!}{\includegraphics{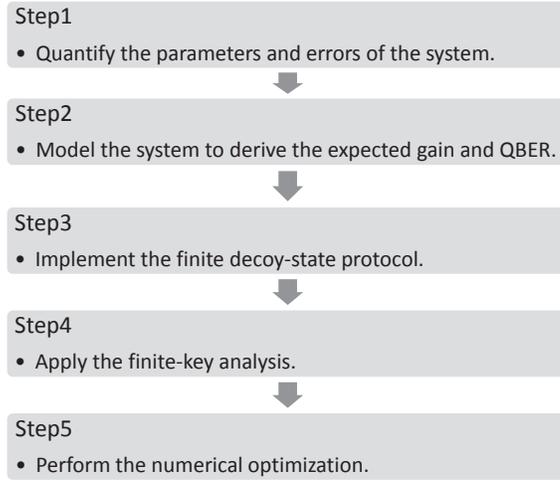}} \caption{\textbf{Framework to choose the optimal intensities.} Step 1 is to quantify the parameters and errors of the system (see Table~\ref{Tab:exp:parameters} for some representative values). Step 2 is to model the system, \ie, derive the gain and QBER by incorporating the practical error sources (see~\ref{App:QandE} for the case of a polarization based system). Step 3 is to implement the finite decoy-state protocol (see Section~\ref{Sec:finitedecoy}). Step 4 is to apply the finite-key analysis~\cite{marcos:finite:2013}. Step 5 is to perform the numerical optimization to get the optimal intensities as well as other parameters such as the optimal selection for the probabilities of different intensity settings.} \label{Fig:flowchart}
\end{figure}
In this section, we develop a general framework to choose the optimal intensity values for the signal and decoy states. This framework is shown in Fig.~\ref{Fig:flowchart}, and is composed of five steps.
\begin{enumerate}
  \item Quantify the parameters and errors of the system. For simulation purposes, we will consider the parameters shown in Table~\ref{Tab:exp:parameters}.

  \item Model the system using the techniques presented in Section~\ref{Sec:errors}. A complete model for a polarization-encoding MDI-QKD can be found in~\ref{App:QandE}.

  \item Implement the finite decoy-state protocol. For this, we will consider the analytical method with two decoy states introduced in Section~\ref{Sec:finitedecoy}. In the simulation, for the weakest decoy state $\omega$, we set its minimum value at $5\times10^{-4}$ (per pulse)~\cite{intensityw}.

  \item Apply the finite-key analysis. Here, we employ the rigorous finite-key analysis of~\cite{marcos:finite:2013} and consider a total number of signals $N$=$10^{14}$~\cite{signaldistribution} together with a security bound of $\epsilon$=$10^{-10}$.

  \item Perform the numerical optimization. In our simulation, we use a MATLAB program to maximize the secure key rate and thus obtain the optimal parameters under different channel transmittances.
\end{enumerate}

\begin{figure}[!t]
\centering
\resizebox{7cm}{!}{\includegraphics{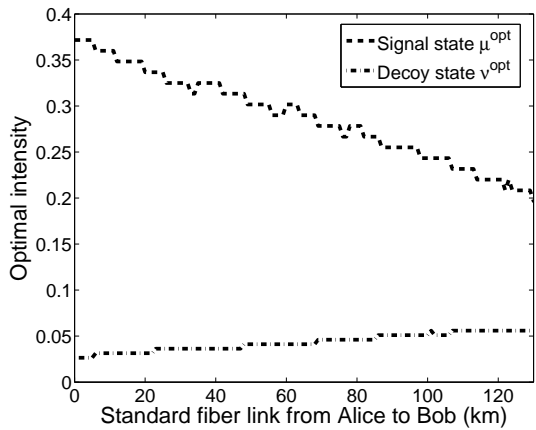}} \caption{\textbf{Optimal intensities.} These intensity values are obtained by numerically maximizing the key rate in the finite-key case. They correspond to the situation where $\mu_a^{opt}$=$\mu_b^{opt}$=$\mu^{opt}$ and $\nu_a^{opt}$=$\nu_b^{opt}$=$\nu^{opt}$. The other decoy state $\omega$ is optimized at its minimum value, \ie, $\omega_a^{opt}$=$\omega_b^{opt}$=$5\times10^{-4}$. Here we use the method described in Section~\ref{Sec:finitedecoy} for the finite decoy-state protocol and that of~\cite{marcos:finite:2013} for the finite-key analysis. For this, we consider a total number of signals $N$=$10^{14}$ together with a security bound of $\epsilon$=$10^{-10}$. The non-smooth behaviors in the figure are mainly due to the lack of numerical accuracy.} \label{Fig:Optimal:intensity}
\end{figure}
Based on this framework, the optimal intensities that maximize the key rate at different transmission distances are shown in Fig.~\ref{Fig:Optimal:intensity}. Notice also that our approach has already been applied to the experimental demonstration reported in~\cite{zhiyuan:experiment:2013}, where the polarization misalignment is around 0.7\% and the total mode mismatch is below 2\%. Owing to the low operation rate there, the value of $\omega$ is set to 0.01. The optimal intensities in this scenario are $\mu_a^{opt}$=$\mu_b^{opt}$$\approx$0.3 and $\nu_a^{opt}$=$\nu_b^{opt}$$\approx$0.1.

%%%%%%%%%%%%%%%%%%%%%%%%%%%%%%%%%%%%%%%%%%%%%%%%%%%%%%%%%%%%%%%%%%%%%%%%%%%%%%%%%%%%%%%%
\section{Asymmetric MDI-QKD} \label{Sec:asymmetric}
 A schematic diagram of the asymmetric MDI-QKD is shown in Fig.~\ref{Fig:general}. Note that this asymmetric scenario appeared naturally in a recent field-test experiment performed in Calgary~\cite{Tittel:2012:MDI:exp}. Another concrete illustration can be found at the Tokyo QKD network~\cite{sasaki2011field}, in which the asymmetric case occurs if Koganei-1 (Alice) and Koganei-3 (Bob) use Koganei-2 (Charles) as the quantum relay to perform MDI-QKD, where the two fiber links are respectively 90 km and 1 km. Here we define a parameter $x$ to quantify the ratio of the two channel transmittances, \ie, $x=t_{a}/t_{b}$. In the Calgary's system, $x$=0.752, while in the Tokyo QKD network $x$=0.017.

\subsection{Problem identification}
The main question here is how to choose the optimal intensities in this asymmetric situation. In the asymptotic case, these optimal intensities refer to the two signal states $\mu_{a}^{opt}$ and $\mu_{b}^{opt}$. Let us discuss two possible options.

The first option is to choose $\mu_{a}^{opt}$=$\mu_{b}^{opt}$ with both in $O(1)$. If we ignore the system imperfections such as background counts and other practical errors, the error rates ($e^{1,1}_{X}$ and $E_{Z}$ in Eq.~\ref{Eqn:Key:formula}) will be zero, while $P^{1,1}_{Z}Y^{1,1}_{Z}$ can be maximized with $\mu_{a}^{opt}$=$\mu_{b}^{opt}$=1 (see~\ref{Sec:Y11e11} for the details). However, in practice, it is inevitable to have some practical errors such as the polarization misalignment discussed above. A relatively large intensity in the short channel will significantly increase the QBER due to the misalignment. Moreover, owing to the intensity mismatch on Charles's side ($\mu_{a}^{opt}t_{a}$$\neq$$\mu_{b}^{opt}t_{b}$), the quantum interference known as the HOM dip, will be imperfect. As a consequence, this option leads to a relatively large QBER, which decreases the key rate due to the cost of error correction.

To minimize the QBER, a second option is to choose $\mu_{a}^{opt}t_{a}$=$\mu_{b}^{opt}t_{b}$ regardless of $x$. We denote this situation as the \emph{symmetric choice} (indicated by Symmetry in Fig.~\ref{Fig:advantage:key} and Table~\ref{Tab:Opt:Parameters}). An equivalent implementation scheme for this option is to add a tailored length of fiber in the local station of the sender with the short channel transmission (\ie, Bob in Fig.~\ref{Fig:general}) in order to balance the two channel transmittances. In fact, such a scheme was recently implemented in a proof-of-principle MDI-QKD experiment \cite{Tittel:2012:MDI:exp}. However, when $x$ is far from 1, to satisfy $\mu_{a}t_{a}$=$\mu_{b}t_{b}$, either $\mu_{a}$ or $\mu_{b}$ needs to be relatively small. Hence, we cannot derive good bounds for $P^{1,1}_{Z}Y^{1,1}_{Z}$ and $e^{1,1}_{X}$. In particular, the increase of $e^{1,1}_{X}$ results in the decrease of the key rate due to the cost of privacy amplification.

In summary, we find that both of the above two options are sub-optimal. We present the optimal choice below.

\subsection{Summary of results}
\begin{figure}[!t]
\centering
\resizebox{7.5cm}{!}{\includegraphics{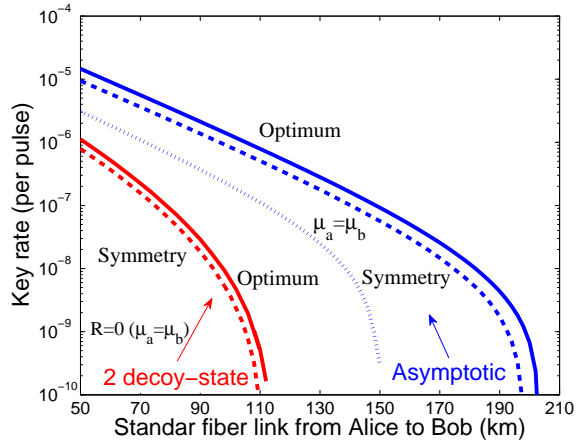}} \caption{\textbf{Key rate comparision with 50km channel mismatch.} We assume a fixed channel mismatch, $x$=0.1 ($L_{ac}$-$L_{bc}$=50km). In the symmetric choice (Symmetry in figure), we set $\mu_{a}t_{a}$=$\mu_{b}t_{b}$, while in the optimal choice (Optimum in figure), we non-trivially determine the optimal intensities by numerical simulation. The red curves are evaluated by the two decoy-state protocol (Section~\ref{Sec:finitedecoy}) combined with the finite-key analysis of~\cite{marcos:finite:2013}. Note that in each curve, all the intensities of the signal and decoy states are optimized by maximizing the key rate. On average, the key rate with the optimal choice is around 80\% larger than that with the symmetric choice in both asymptotic and two decoy-state cases.} \label{Fig:advantage:key}
\end{figure}

\begin{table}[hbt]
\centering
\begin{tabular}{c|cccc|cccc}
\hline
           & \multicolumn{4}{c|}{Asymptotic case} & \multicolumn{4}{c}{Two decoy-state case} \\
\hline
Parameters & \multicolumn{2}{c|}{Symmetry} & \multicolumn{2}{c|}{Optimum}  & \multicolumn{2}{c|}{Symmetry} & \multicolumn{2}{c}{Optimum}\\
\hline
$x=0.1$  & $\mu_{a}$ & $\mu_{b}$ & $\mu_{a}^{opt}$ & $\mu_{b}^{opt}$ & $\mu_{a}$ & $\mu_{b}$ & $\mu_{a}^{opt}$ & $\mu_{b}^{opt}$ \\
\hline
$L_{bc}$=0km  & 0.75 & 0.08 & 0.60 & 0.15 & 0.49 & 0.05  & 0.46 & 0.08  \\
\hline
$L_{bc}$=10km & 0.75 & 0.08 & 0.60 & 0.15 & 0.46 & 0.05  & 0.41 & 0.07  \\
\hline
$L_{bc}$=20km & 0.75 & 0.08 & 0.60 & 0.15 & 0.41 & 0.03  & 0.38 & 0.06  \\
\hline
\end{tabular}
\caption{\textbf{Optimal intensities of an asymmetric MDI-QKD system.} The channel mismatch is fixed at $x$=0.1 and thus $L_{ac}$=\{50km, 60km, 70km\}. In the asymptotic case, the ratio $\mu_{a}^{opt}$/$\mu_{b}^{opt}$ for the optimal choice is around 4. In the symmetric choice, this ratio is 10. The parameters $\mu_{a}^{opt}$ and $\mu_{b}^{opt}$ are fixed regardless of $t_{a}$ and $t_{b}$ (see Theorem~\ref{theorem1} in~\ref{Sec:asymptotic:estimation}). In the two decoy-state case (with different $L_{bc}$), the optimal intensities for the decoy state $\nu$ are about \{$\nu_{a}$=0.10, $\nu_{b}$=0.01\} for the symmetric choice and \{$\nu_{a}^{opt}$=0.07, $\nu_{b}^{opt}$=0.01\} for the optimal choice. The optimal value for the weakest decoy state is $\omega_a^{opt}$=$\omega_b^{opt}$=$5\times10^{-4}$ for both choices. From the intensity values in this table, we find that the optimal choice for $\mu_{a}$ and $\mu_{b}$ does not always satisfy $\mu_{a}^{opt}t_{a}$=$\mu_{b}^{opt}t_{b}$, but the ratio $\mu_{a}^{opt}t_{a}/\mu_{b}^{opt}t_{b}$ is near 1. Also, in the asymptotic case, $\mu_{a}^{opt}$ and $\mu_{b}^{opt}$ are \emph{only} determined by $x$ instead of $t_{a}$ or $t_{b}$. } \label{Tab:Opt:Parameters}
\end{table}

The \emph{optimal choice} (indicated by Optimum in Fig.~\ref{Fig:advantage:key} and Table~\ref{Tab:Opt:Parameters}) that maximizes the key rate can be determined from numerical optimizations. Here we perform such optimizations and also analyze the properties of asymmetric MDI-QKD. Our main results are:

%also analytically and numerically optimize its key rate  a non-trivial choice between the condition of $\mu_{a}$=$\mu_{b}$ and that of $\mu_{a}t_{a}$=$\mu_{b}t_{b}$. This non-trivial choice is

\begin{enumerate}
  \item In the asymptotic case, the optimal choice for $\mu_{a}$ and $\mu_{b}$ does not always satisfy $\mu_{a}^{opt}t_{a}$=$\mu_{b}^{opt}t_{b}$, but the ratio $\mu_{a}^{opt}t_{a}/\mu_{b}^{opt}t_{b}$ is near 1. For $x$$<$1, $\mu_{a}^{opt}t_{a}/\mu_{b}^{opt}t_{b}$$\in$[0.3, 1); for $x$$\geq$1, $\mu_{a}^{opt}t_{a}/\mu_{b}^{opt}t_{b}$$\in$[1, 3.5]; This result can be seen from Fig.~\ref{Simu:mumu:withdark}). In the practical case with the two decoy-state protocol and finite-key analysis, \{$\mu_{a}^{opt}$, $\mu_{b}^{opt}$\} and \{$\nu_{a}^{opt}$, $\nu_{b}^{opt}$\} satisfy a similar condition with the ratio $\mu_{a}^{opt}t_{a}/\mu_{b}^{opt}t_{b}$ (or $\nu_{a}^{opt}t_{a}/\nu_{b}^{opt}t_{b}$) near 1, while \{$\omega_{a}^{opt}$, $\omega_{b}^{opt}$\} are optimized at their smallest value. See Table~\ref{Tab:Opt:Parameters} for further details.

  \item In an asymmetric system with $x$=0.1 (50 km length difference for two standard fiber links), the advantage of the optimal choice is shown in Fig.~\ref{Fig:advantage:key}, where the key rate with the optimal choice is around 80\% larger than that with the symmetric choice in both asymptotic and practical cases~\cite{differentN}. We remark that when $x$ is far from 1, this advantage is more significant. For instance, with $x$=0.01 (100 km length difference), the key rate with the optimal choice is about 150\% larger than that with the symmetric choice.

  \item In the asymptotic case, at a short distance where background counts can be ignored: $\mu_{a}^{opt}$ and $\mu_{b}^{opt}$ are \emph{only} determined by $x$ instead of $t_{a}$ or $t_{b}$ (see the optimal intensities in Table~\ref{Tab:Opt:Parameters} and Theorem~\ref{theorem1} in~\ref{App:asymptotic}); assuming a fixed $x$, $\mu_{a}^{opt}$ and $\mu_{b}^{opt}$ can be analytically derived and the optimal key rate is quadratically proportional to $t_{b}$ (see~\ref{App:asymptotic}).
\end{enumerate}

Finally, notice that the channel transmittance ratio in Calgary's asymmetric system is near 1 ($x$=0.752), hence the optimal choice can slightly improve the key rate compared to the symmetric choice (around 2\% improvement). However, in Tokyo's asymmetric system ($x$=0.017), the optimal choice can significantly improve the key rate by over 130\%.

\section{Discussion and Conclusion}\label{Sec:conclusion}
A key assumption in MDI-QKD~\cite{Lo:MDIQKD} is that Alice and Bob trust their devices for the state preparation, \ie, they can generate ideal quantum states in the BB84 protocol. One approach to remove this assumption is to quantify the imperfections in the state preparation part and thus include them into the security proofs~\cite{tamaki2012phase}. We believe that this assumption is practical because Alice's and Bob's quantum states are prepared by themselves and thus can be experimentally verified in a fully protected laboratory environment outside of Eve's interference. For instance, based on an earlier proposal~\cite{braunstein2012side}, C. C. W. Lim \emph{et al.} have introduced another interesting scheme~\cite{lim2012quantum} in which each of Alice and Bob uses an entangled photon source (instead of WCPs) and quantifies the state-preparation imperfections via random sampling. That is, Alice and Bob randomly sample parts of their prepared states and perform a local Bell test on these samples. Such a scheme is very promising, as it is in principle a fully device-independent approach. It can be applied in short-distance communications.

In conclusion, we have presented an analysis for practical aspects of MDI-QKD. To understand the physical origin of the QBER, we have investigated various practical error sources by developing a general system model. In a polarization-encoding MDI-QKD system, polarization misalignment is the major source contributing to the QBER. Hence, in practice, an efficient polarization management scheme such as polarization feedback control~\cite{da2012proof, polarizationcontrol} can significantly improve the polarization stabilization and thus generate a higher key rate. We have also discussed a simple analytical method for the finite decoy-state analysis, which can be directly used by experimentalists to demonstrate MDI-QKD. In addition, by combining the system model with the finite decoy-state method, we have presented a general framework for the optimal intensities of the signal and decoy states. Furthermore, we have studied the properties of the asymmetric MDI-QKD protocol and discussed how to optimize its performance. Our work is relevant to both QKD and general experiments on quantum interference.

%%%%%%%%%%%%%%%%%%%%%%%%%%%%%%%%%%%%%%%%%%%%%%%%%%%%%%%%%%%%%%%%%%%%
%% Acknowledgments
\section*{Acknowledgments}
We thank W.~Cui, S. Gao, L.~Qian for enlightening discussions and V.~Burenkov, Z.~Liao, P.~Roztocki for comments on the presentation of the paper. Support from funding agencies NSERC, the CRC program, European Regional Development Fund (ERDF), and the Galician Regional Government (projects CN2012/279 and CN 2012/260, ``Consolidation of Research Units: AtlantTIC") is gratefully acknowledged. F. Xu would like to thank the Paul Biringer Graduate Scholarship for financial support.

%\newpage
\begin{appendix}

\section{Other practical errors} \label{App:othererrors}
Here, we discuss other practical error sources and show that their contribution to the QBER is not very significant in a practical MDI-QKD system. For this reason, they are ignored in our simulations.

%%%%%%%%%%%%%%%%%%%%%%%%%%%%%%%%%%%%%%%%%%%%%%%%%%%%%%%%%%%%%%%%%%%%%%%%%%%%%%%%%%%%%%%%
\subsection{Intensity fluctuations at the source} \label{intensity:mismatch:section}
The intensity fluctuations of the signal and decoy states at the source are relatively small ($\sim$ 0.1 dB)~\cite{rosenberg2009practical}. Additionally, Alice and Bob can in principle locally and precisely quantify their own intensities. Therefore, this error source can be mostly ignored in the theoretical model that analyzes the performance of practical MDI-QKD (but one could easily include it in the analysis).

%%%%%%%%%%%%%%%%%%%%%%%%%%%%%%%%%%%%%%%%%%%%%%%%%%%%%%%%%%%%%%%%%%%%%%%%%%%%%%%%%%%%%%%%
\subsection{Threshold detector with background counts}
\begin{figure}[!t]
\centering
\resizebox{6.7cm}{!}{\includegraphics{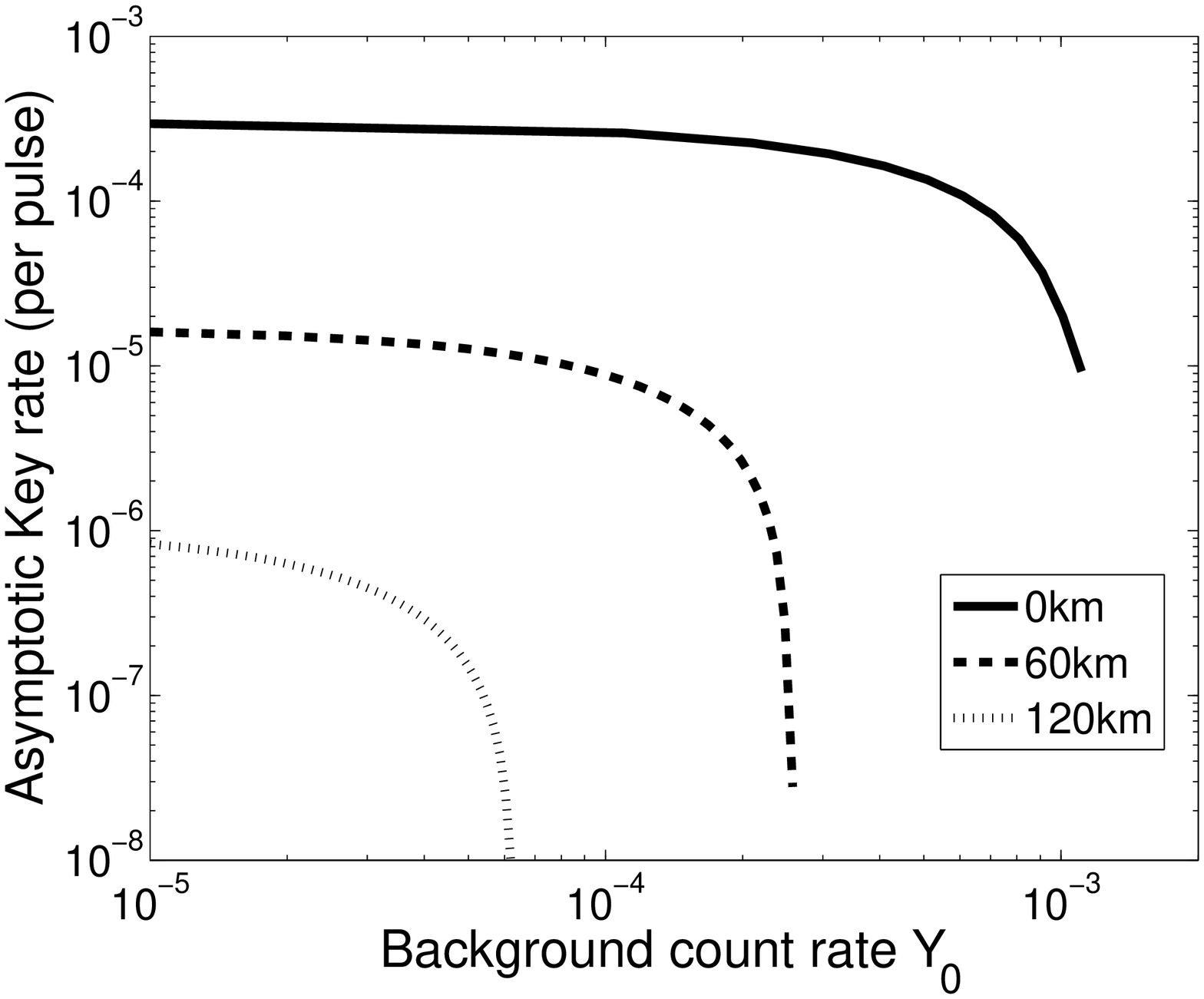}} \caption{\textbf{Background counts tolerance.} Following the model discussed in Section~\ref{App:QandE}, we simulate the asymptotic key rates at different background count rates. MDI-QKD is robust to background counts.} \label{detector:darkcounts}
\end{figure}
The threshold single photon detector (SPD) can be modeled by a beam splitter with $\eta_{d}$ transmission and (1-$\eta_{d}$) reflection. The transmission part is followed by a unity efficiency detector, while the reflection part is discarded. $\eta_{d}$ is defined as the detector efficiency. Background counts can be treated to be independent of the incoming signals. For simplicity, the system model discussed in Section~\ref{App:QandE} assumes that the four SPDs (see Fig.~\ref{Fig:model}) are identical and have a detection efficiency $\eta_{d}$ and a background rate $Y_0$. Note, however, that if this condition is not satisfied (\ie, there is some detection efficiency mismatch) our system model can be adapted to take care also of this case.

All the simulations reported in the main text already consider a background rate of $Y_{0}$=$6.02\times10^{-6}$ (see Table~\ref{Tab:exp:parameters}). Fig.~\ref{detector:darkcounts} simulates more general cases of the asymptotic key rates at different background count rates. At 0 km, the MDI-QKD system can tolerate up to $10^{-3}$ (per pulse) background counts.
%%%%%%%%%%%%%%%%%%%%%%%%%%%%%%%%%%%%%%%%%%%%%%%%%%%%%%%%%%%%%%%%%%%%%%%%%%%%%%%%%%%%%%%%
\subsection{Beam splitter ratio}
\begin{figure}[!t]
\centering
\resizebox{7cm}{!}{\includegraphics{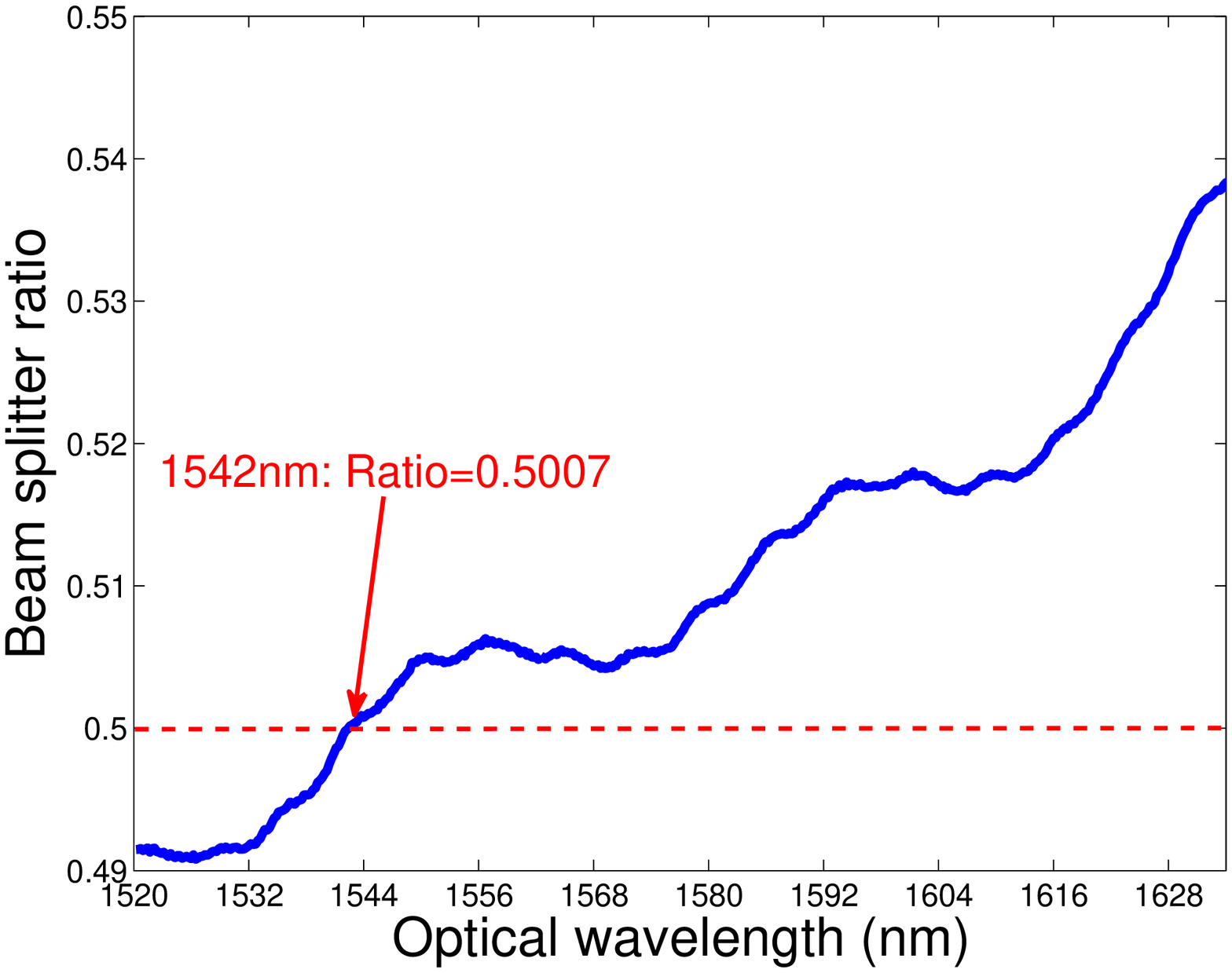}} \caption{\textbf{Wavelength dependence of a fiber-based beam splitter.} If the laser wavelength is 1542 nm~\cite{zhiyuan:experiment:2013}, the beam splitter ratio is 0.5007, which introduces negligible QBER (below 0.01\%) in a typical MDI-QKD system.} \label{Fig:BSratio}
\end{figure}

In practice, for telecom wavelengths, the asymmetry of the beam splitter (BS) (\ie, not 50:50) is usually small. For instance, the wavelength dependence of the fiber-based BS in our lab (Newport-1310\/1550-50\/50 fiber coupler) is experimentally quantified in Fig.~\ref{Fig:BSratio}. If the laser wavelength is 1542 nm~\cite{zhiyuan:experiment:2013}, the BS ratio is 0.5007, which introduces negligible QBER (below 0.01\%) in a MDI-QKD system. Hence, this error source can also be ignored in the theoretical model of MDI-QKD.

\section{System model -- analytical key rate} \label{App:QandE}
In this section, we discuss an analytical method to model a polarization-encoding MDI-QKD system. That is, we calculate $Y_{Z}^{1,1}$, $e_{X}^{1,1}$, $Q_Z$ and $E_Z$ and thus estimate the expected key rate from Eq.~(\ref{Eqn:Key:formula}).

To simplify our calculation, we make two assumptions about the practical error sources: a) since most practical error sources do not contribute significantly to the system performance, we only consider the polarization misalignment $e_d$, the background count rate $Y_0$ and the detector efficiency $\eta_d$; b) for the model of the polarization misalignment, we consider only two unitary operators, $U_{1}$ and $U_{2}$, to represent respectively the polarization misalignment of Alice's and Bob's channel transmission, \ie, set $U_{3}$=$I$ in the generic model of Sec.~\ref{Sec:polarization}. For simplicity, a more rigorous derivation with $U_3$ $\neq I$ is not shown here, but it can be easily completed following our procedures discussed below.

\subsection{$Y_{Z}^{1,1}$ and $e_{X}^{1,1}$} \label{Sec:Y11e11}
In the asymptotic case, we assume that $Y_{Z}^{1,1}$ and $e_{X}^{1,1}$ in Eq.~(\ref{Eqn:Key:formula}) can be perfectly estimated with an infinite number of signals and decoy states. Thus, they are given by
\begin{eqnarray} \label{Q11e11:nonideal:final}
    e_{X}^{1,1}=\frac{1}{2}-\ \frac{t_{a}t_{b}\eta_{d}^{2}(1-e_{d})^{2}(1-Y_{0})^{2}}{4Y_{X}^{1,1}}, \\ \nonumber
    Y_{Z}^{1,1}=(1-Y_{0})^{2}[4Y_{0}^{2}(1-t_{a}\eta_{d})(1-t_{b}\eta_{d})\\ \nonumber
               +2Y_{0}(t_{a}\eta_{d}+t_{b}\eta_{d}-\frac{3t_{a}t_{b}\eta_{d}^{2}}{2})+\frac{t_{a}t_{b}\eta_{d}^{2}}{2}],
\end{eqnarray}
where $Y_{X}^{1,1}$=$Y_{Z}^{1,1}$. Importantly, we can see that ignoring the imperfections of polarization misalignment and background counts (\ie, $e_d=0$, $Y_0=0$), $e^{1,1}_{X}$ is zero, while $P^{1,1}_{Z}Y^{1,1}_{Z}$ ($P^{1,1}_{Z}$=$\mu_a\mu_b e^{-(\mu_a+\mu_b)}$) can be maximized with $\mu_{a}$=$\mu_{b}$. Thus, the optimal choice of intensities is $\mu_{a}$=$\mu_{b}$=1. However, in practice, it is inevitable to have certain practical errors, which result in this optimal choice being a \emph{function} of the values of practical errors.

\subsection{$Q_{Z}$ and $E_{Z}$}
Now, let us calculate $Q_{Z}$ and $E_{Z}$, which are eventually given by Eq.~(\ref{Qrect:Erect}). To further simplify our discussion, we use \{horizonal,vertical,45-degree,135-degree\} to represent the BB84 polarization states. Also, \{$HH, HV, ++, +-$\} will denote Alice's and Bob's encoding modes. We define the following notations:

\begin{eqnarray} \label{semiideal:notations} \nonumber
    \gamma_{a}=\sqrt{\mu_{a}t_{a}\eta_{d}}, \gamma_{b}=\sqrt{\mu_{b}t_{b}\eta_{d}},  \\ \nonumber
    \beta=\gamma_{a}\gamma_{b}, \gamma=\gamma_{a}^{2}+\gamma_{b}^{2}, \\ \nonumber
    \lambda=\gamma_{a}\gamma_{b}\sqrt{e_{d1}(1-e_{d1})}, \omega=\gamma_{a}^{2}+e_{d1}(\gamma_{b}^{2}-\gamma_{a}^{2}).
\end{eqnarray}

\subsubsection{Derivation of $Q_{Z}^{HH}$} \label{Sec:QHH}
First, both Alice and Bob encode their states in the H mode (symmetric to V mode). We assume that $U_{1}$ and $U_{2}$ (see Eq.~(\ref{Unitary})) rotate the polarization in the \emph{same} direction, i.e. $\theta_{1}\theta_{2}>0$. The discussion regarding rotation in the opposite direction (i.e. $\theta_{1}\theta_{2}<0$) is in~\ref{QHH:oppositeangle}.

In Charles's lab, after the BS and PBS (see Fig.~\ref{Fig:model}), the optical intensities received by each SPD are given by
\begin{eqnarray} \label{QHH:meanphotonnumbers:HH}
    D_{ch}: |A|^{2}=\frac{(1-e_{d1})(\gamma_{a}^{2}+\gamma_{b}^{2})-2\gamma_{a}\gamma_{b}\cos(\phi)(1-e_{d1})}{2}, \\ \nonumber
    D_{dh}: |C|^{2}=\frac{(1-e_{d1})(\gamma_{a}^{2}+\gamma_{b}^{2})+2\gamma_{a}\gamma_{b}\cos(\phi)(1-e_{d1})}{2},\\ \nonumber
    D_{cv}: |B|^{2}=\frac{e_{d1}(\gamma_{a}^{2}+\gamma_{b}^{2})-2\gamma_{a}\gamma_{b}\cos(\phi)e_{d1}}{2}, \\ \nonumber
    D_{dv}: |D|^{2}=\frac{e_{d1}(\gamma_{a}^{2}+\gamma_{b}^{2})+2\gamma_{a}\gamma_{b}\cos(\phi)e_{d1}}{2},
\end{eqnarray}
where $\phi$ denotes the relative phase between Alice's and Bob's weak coherent states. Thus, the detection probability of each threshold SPD is
\begin{equation} \label{QHH:detection:prob}
  P_{V}=1-(1-Y_{0})e^{-|V|^{2}},
\end{equation}
where $V=A,B,C,D$. Then, the coincident counts are
\begin{eqnarray} \label{QHH:model:final} \nonumber
    Q_{Z}^{HH,\psi^{+}}=2P_{A}P_{B}(1-P_{C})(1-P_{D}), \\ \nonumber
    Q_{Z}^{HH,\psi^{-}}=2P_{A}P_{D}(1-P_{B})(1-P_{C}),
\end{eqnarray}
where $Q_{Z}^{HH,\psi^{+}}$ and $Q_{Z}^{HH,\psi^{-}}$ denote, respectively, the probability of the projection on the Triplet $|\psi^{+}\rangle$= $\frac{1}{\sqrt{2}}(|H,V\rangle+|V,H\rangle$) and the Singlet $|\psi^{-}\rangle$= $\frac{1}{\sqrt{2}}(|H,V\rangle-|V,H\rangle$). Here from Fig.~\ref{Fig:model}, Triplet means the coincident detections of \{ch \& cv\} or \{dh \& dv\}; Singlet means the coincident detections of \{ch \& dv\} or \{cv \& dh\}. After averaging over the relative phase $\phi$ (integration over [0,2$\pi$]), we have
\begin{eqnarray}
Q_{Z}^{HH,\psi^{+}}&=2e^{-\frac{\gamma}{2}}(1-Y_{0})^{2}[I_{0}(\beta)+(1-Y_{0})^{2}e^{-\frac{\gamma}{2}} \\ \nonumber
    &-(1-Y_{0})e^{-\frac{\gamma(1-e_{d1})}{2}}I_{0}(e_{d1}\beta)-(1-Y_{0})e^{-\frac{\gamma e_{d1}}{2}}I_{0}(\beta-e_{d1}\beta)], \\ \nonumber
Q_{Z}^{HH,\psi^{-}}&=2e^{-\frac{\gamma}{2}}(1-Y_{0})^{2}[I_{0}(\beta-2\beta e_{d1})+(1-Y_{0})^{2}e^{-\frac{\gamma}{2}} \\ \nonumber
    &-(1-Y_{0})e^{-\frac{\gamma(1-e_{d1})}{2}}I_{0}(e_{d1}\beta)-(1-Y_{0})e^{-\frac{\gamma e_{d1}}{2}}I_{0}(\beta-e_{d1}\beta)],
\end{eqnarray} \label{QHH:final}
where $I_{0}(\cdot)$ is the modified Bessel function. Therefore, $Q_{Z}^{HH}$ is given by
\begin{equation} \label{QHH:finalfinal}
Q_{Z}^{HH}=Q_{Z}^{HH,\psi^{+}}+Q_{Z}^{HH,\psi^{-}}.
\end{equation}

Here, to simplify Eq.~(\ref{QHH:final}), we ignore background counts, \ie, $Y_{0}=0$, and use a 2nd order approximation (as both $\beta$ and $\gamma$ are typically on the order of $0.01$) such that
\begin{eqnarray} \label{general:estimation} \nonumber
    I_{0}(\beta)=1+\frac{\beta^{2}}{4}+O(\beta^{4}), \\ \nonumber
    e^{\gamma}=1+\gamma+\frac{\gamma^{2}}{2}+O(\gamma^{3}),
\end{eqnarray}
then, Eq.~(\ref{QHH:final}) can be estimated as
\begin{eqnarray} \label{QHH:final:estimation:two}
    Q_{Z}^{HH,\psi^{+}}=\frac{\gamma^{2}e_{d1}(1-e_{d1})}{2}+\beta^{2}e_{d1}(1-e_{d1}), \\ \nonumber
    Q_{Z}^{HH,\psi^{-}}=\frac{\gamma^{2}e_{d1}(1-e_{d1})}{2}-\beta^{2}e_{d1}(1-e_{d1}),
\end{eqnarray}
and $Q_{Z}^{HH}$ is given by
\begin{equation} \label{QHH:final:estimation}\nonumber
   Q_{Z}^{HH}=\gamma^{2}e_{d1}(1-e_{d1}).
\end{equation}

\subsubsection{Derivation of $Q_{Z}^{HV}$} \label{Sec:QHV}
Alice (Bob) encodes her (his) state in the H (V) mode (symmetric to V (H)). We also assume $\theta_{1}\theta_{2}>0$. At Charles's side, the optical intensities received by each SPD are given by
\begin{eqnarray} \label{QHV:meanphotonnumbers} \nonumber
    |A'|^{2}=\frac{(1-e_{d1})\gamma_{a}^{2}+e_{d1}\gamma_{b}^{2}-2\lambda\cos(\phi)}{2}, \\ \nonumber
    |B'|^{2}=\frac{e_{d1}\gamma_{a}^{2}+(1-e_{d1})\gamma_{b}^{2}-2\lambda\cos(\phi)}{2}, \\ \nonumber
    |C'|^{2}=\frac{(1-e_{d1})\gamma_{a}^{2}+e_{d1}\gamma_{b}^{2}+2\lambda\cos(\phi)}{2}, \\ \nonumber
    |D'|^{2}=\frac{e_{d1}\gamma_{a}^{2}+(1-e_{d1})\gamma_{b}^{2}+2\lambda\cos(\phi)}{2}.
\end{eqnarray}

The detection probability of each SPD is described by Eq.~(\ref{QHH:detection:prob}). $Q_{Z}^{HV,\psi^{+}}$ and $Q_{Z}^{HV,\psi^{-}}$ can be calculated similarly to Eq.~(\ref{QHH:model:final}). After averaging over $\phi$, the results are
\begin{eqnarray} \label{QHV:final}
    Q_{Z}^{HV,\psi^{+}}&=2e^{-\frac{\gamma}{2}}(1-Y_{0})^{2}[I_{0}(2\lambda)+(1-Y_{0})^{2}e^{-\frac{\gamma}{2}}\\ \nonumber
    &-(1-Y_{0})e^{-\frac{\omega}{2}}I_{0}(\lambda)-(1-Y_{0})e^{-\frac{\gamma-\omega}{2}}I_{0}(\lambda)],\\ \nonumber
    Q_{Z}^{HV,\psi^{-}}&=2e^{-\frac{\gamma}{2}}(1-Y_{0})^{2}[1+(1-Y_{0})^{2}e^{-\frac{\gamma}{2}}\\ \nonumber
    &-(1-Y_{0})e^{-\frac{\omega}{2}}I_{0}(\lambda)-(1-Y_{0})e^{-\frac{\gamma-\omega}{2}}I_{0}(\lambda)].
\end{eqnarray}

Therefore, $Q_{Z}^{HV}$ is given by
\begin{equation} \label{QHV:finalfinal}
Q_{Z}^{HV}=Q_{Z}^{HV,\psi^{+}}+Q_{Z}^{HV,\psi^{-}}.
\end{equation}

To simplify Eq.~(\ref{QHV:final}) we once again ignore the background counts and take a 2nd order approximation. Eq.~(\ref{QHV:final}) can be estimated as
\begin{eqnarray} \label{QHV:final:estimation:two}
    Q_{Z}^{HV,\psi^{+}}=\frac{\omega(\gamma-\omega)}{2}+\lambda^{2}, \\ \nonumber
    Q_{Z}^{HV,\psi^{-}}=\frac{\omega(\gamma-\omega)}{2}-\lambda^{2},
\end{eqnarray}
and $Q_{Z}^{HV}$ is given by
\begin{equation} \label{QHV:final:estimation} \nonumber
   Q_{Z}^{HV}=\omega(\gamma-\omega).
\end{equation}

\subsubsection{Derivation of $Q_{Z}$ and $E_{Z}$}
Finally, $Q_{Z}$ and $E_{Z}$ can be expressed as
\begin{eqnarray} \label{Qrect:Erect}
    Q_{Z}=\frac{Q_{Z}^{HH}+Q_{Z}^{HV}}{2},\\ \nonumber
    E_{Z}=\frac{Q_{Z}^{HH}}{Q_{Z}^{HH}+Q_{Z}^{HV}},
\end{eqnarray}
where the different terms on the r.h.s. of this equation are given by Eqs.~(\ref{QHH:final}, \ref{QHH:finalfinal}, \ref{QHV:final}, \ref{QHV:finalfinal}). Therefore, together with Eq.~(\ref{Q11e11:nonideal:final}), we could derive the analytical key rate of Eq.~(\ref{Eqn:Key:formula}).

If we ignore background counts and take the 2nd order approximation from Eqs.~(\ref{QHH:final:estimation:two}, \ref{QHV:final:estimation:two}), $Q_{Z}$ and $E_{Z}$ can be written as
\begin{eqnarray} \label{Qrect:Erect:final:estimation}
    Q_{Z}=\frac{\beta^{2}+e_{d}(1-\frac{e_{d}}{2})(\gamma^{2}-2\beta^{2})}{2}, \\ \nonumber
    E_{Z}=\frac{\gamma^{2}e_{d}(1-\frac{e_{d}}{2})}{4Q_{Z}}.
\end{eqnarray}

\subsubsection{$Q_{Z}$ and $E_{Z}$ with opposite rotation angle} \label{QHH:oppositeangle}
When $U_{1}$ and $U_{2}$ rotate the polarization in the \emph{opposite} direction, \ie, $\theta_{1}\theta_{2}<0$, Eq.~(\ref{QHH:meanphotonnumbers:HH}) changes to
\begin{eqnarray} \label{QHH:opposite:meanphotonnumbers:HH}\nonumber
   |A|^{2}=\frac{(1-e_{d1})(\gamma_{a}^{2}+\gamma_{b}^{2})-2\gamma_{a}\gamma_{b}\cos(\phi)(1-e_{d1})}{2}, \\ \nonumber
   |C|^{2}=\frac{(1-e_{d1})(\gamma_{a}^{2}+\gamma_{b}^{2})+2\gamma_{a}\gamma_{b}\cos(\phi)(1-e_{d1})}{2},\\ \nonumber
   |B|^{2}=\frac{e_{d1}(\gamma_{a}^{2}+\gamma_{b}^{2})+2\gamma_{a}\gamma_{b}\cos(\phi)e_{d1}}{2},\\ \nonumber
   |D|^{2}=\frac{e_{d1}(\gamma_{a}^{2}+\gamma_{b}^{2})-2\gamma_{a}\gamma_{b}\cos(\phi)e_{d1}}{2}.
\end{eqnarray}

After performing similar procedures to those of Section~\ref{Sec:QHH}, Eq.~(\ref{QHH:final:estimation:two}) is altered to
\begin{eqnarray} \label{QHH:opposite:final:estimation:two}
    Q_{Z}^{HH,\psi^{+}}=\frac{\gamma^{2}e_{d1}(1-e_{d1})}{2}-\beta^{2}e_{d1}(1-e_{d1}), \\ \nonumber
    Q_{Z}^{HH,\psi^{-}}=\frac{\gamma^{2}e_{d1}(1-e_{d1})}{2}+\beta^{2}e_{d1}(1-e_{d1}).
\end{eqnarray}

Since the QBER is mainly determined by $Q_{Z}^{HH}$, by comparing Eq.~(\ref{QHH:final:estimation:two}) to (\ref{QHH:opposite:final:estimation:two}), we conclude that
\begin{description}
  \item[$\theta_{1}\theta_{2}>0$] Projection on $|\psi^{+}\rangle$ results in a larger QBER than that on $|\psi^{-}\rangle$
  \item[$\theta_{1}\theta_{2}<0$] Projection on $|\psi^{+}\rangle$ results in a smaller QBER than that on $|\psi^{-}\rangle$
\end{description}

An equivalent analysis can also be applied to $Q_{Z}^{HV}$ following Section~\ref{Sec:QHV}, and thus Eq.~(\ref{QHV:final:estimation:two}) is altered to
\begin{eqnarray} \label{QHV:opposite:final:estimation:two}
    Q_{Z}^{HV,\psi^{+}}=\frac{\omega(\gamma-\omega)}{2}-\lambda^{2}, \\ \nonumber
    Q_{Z}^{HV,\psi^{-}}=\frac{\omega(\gamma-\omega)}{2}+\lambda^{2}.
\end{eqnarray}

Therefore, the key rates of $R^{\psi^{-}}$ (projections on the Triplet) and $R^{\psi^{+}}$ (projections on the Singlet) are \emph{correlated} with the relative direction of the rotation angles, while the overall key rate $R$ ($R=R^{\psi^{-}}+R^{\psi^{+}}$) is \emph{independent} of the relative direction of the rotation angles.

We finally remark that in a practical polarization-encoding MDI-QKD system, the polarization rotation angle of each quantum channel ($\theta_{1}$ or $\theta_{2}$) can be modeled by a Gaussian distribution with a standard deviation of $\theta_{k}^{std}=\arcsin(\sqrt{e_{k}})$ ($k=1, 2$), which means that both $\theta_{1}$ and $\theta_{2}$ (mostly) distribute in the range of [$-3\theta_{k}^{std}$, $3\theta_{k}^{std}$] and the relative direction between them also randomly distributes between $\theta_{1}\theta_{2}>0$ and $\theta_{1}\theta_{2}<0$. Hence, the effect of the polarization misalignment is the same for $R^{\psi^{-}}$ and $R^{\psi^{+}}$, \ie, both $R^{\psi^{-}}$ and $R^{\psi^{+}}$ are independent of the total polarization misalignment. We can experimentally choose to measure either the Singlet or the Triplet by using only two detectors (but sacrificing half of the total key rate), such as in the experiments of~\cite{zhiyuan:experiment:2013, Tittel:2012:MDI:exp}.

%%%%%%%%%%%%%%%%%%%%%%%%%%%%%%%%%%%%%%%%%%%%%%%%%%%%%%%%%%%%%%%%%%%%%%%%%%
\section{Asymmetric MDI-QKD} \label{App:asymptotic}
Here we discuss the properties of a practical asymmetric MDI-QKD system. For this, we derive an analytical expression for the estimated key rate and we optimize the system performance numerically.

\subsection{Estimated key rate} \label{Sec:asymptotic:estimation}
The estimated key rate $R_{est}$ is defined under the condition that background counts can be ignored. Note that this is a reasonable assumption for a short distance transmission.
\begin{theo} \label{theorem1}
$\mu_{a}^{opt}$ and $\mu_{b}^{opt}$ only depend on $x$ rather than on $t_{a}$ or $t_{b}$; Under a fixed $x$, $R_{est}$ is quadratically proportional to $t_{b}$.
\end {theo}

%\begin{proof}
\emph{Proof}: When $Y_{0}$ is ignored, $e_{X}^{1,1}$ and $P^{1,1}Y_{Z}^{1,1}$ are given by (see Eq.~(\ref{Q11e11:nonideal:final}))
\begin{eqnarray} \label{Q11e11:semiideal:final}\nonumber
    e_{X}^{1,1}=e_{d}-\frac{e_{d}^{2}}{2}, \\ \nonumber
    P^{1,1}_{Z}Y_{Z}^{1,1}=\frac{\mu_{a}t_{a}\mu_{b}t_{b}e^{-(\mu_{a}+\mu_{b})}\eta_{d}^{2}}{2}.
\end{eqnarray}
If we take the 2nd order approximation, $Q_{Z}$ and $E_{Z}$ are estimated as (see Eq.~(\ref{Qrect:Erect:final:estimation}))
\begin{eqnarray} \label{Qrect:Erect:final:estimation:xmuamub}
    Q_{Z}=\frac{t_{b}^{2}\eta_{d}^{2}[2x\mu_{a}\mu_{b}+(\mu_{b}^{2}+x^{2}\mu_{a}^{2})(2e_{d}-e_{d}^{2})]}{4}, \\ \nonumber
    E_{Z}=\frac{(\mu_{b}+x\mu_{a})^{2}(2e_{d}-e_{d}^{2})}{2[2x\mu_{a}\mu_{b}+(\mu_{b}^{2}+x^{2}\mu_{a}^{2})(2e_{d}-e_{d}^{2})]}.
\end{eqnarray}
By combining the above two equations with Eq.~(\ref{Eqn:Key:formula}), the overall key rate can be written as
\begin{equation} \label{Key:formula:xmuamub}
    R_{est}=\frac{t_{b}^{2}\eta_{d}^{2}}{2}G(x,\mu_{a},\mu_{b}),
\end{equation}
where $G(x,\mu_{a},\mu_{b})$ has the form
\begin{eqnarray} \label{Key:formula:Gxmuamub}
    G(x,\mu_{a},\mu_{b})&= x\mu_{a}\mu_{b}e^{-(\mu_{a}+\mu_{b})}[1-H_{2}(e_{d}-\frac{e_{d}^{2}}{2})]\\ \nonumber
                        &-\frac{2x\mu_{a}\mu_{b}+(\mu_{b}^{2}+x^{2}\mu_{a}^{2})(2e_{d}-e_{d}^{2})}{2}\times f_e H_2(E_Z),
\end{eqnarray}
where $E_{Z}$ is given by Eq.~(\ref{Qrect:Erect:final:estimation:xmuamub}) and is also a function of $(x,\mu_{a},\mu_{b})$. Therefore, optimizing $R_{est}$ is equivalent to maximizing $G(x,\mu_{a},\mu_{b})$ and the optimal values, $\mu_{a}^{opt}$ and $\mu_{b}^{opt}$, are \emph{only} determined by $x$. Under a fixed $x$, the optimal key rate is quadratically proportional to $t_{b}$. For a given $x$,  the maximization of $G(x,\mu_{a},\mu_{b})$ can be done by calculating the derivatives over $\mu_{a}$ and $\mu_{b}$ and verified using the Jacobian matrix.
%\end{proof}

%%%%%%%%%%%%%%%%%%%%%%%%%%%%%%%%%%%%%%%%%%%%%%%%%%%%%%%%%%%%%%%%%%%%%%%%%%%%%%%%%%%%%%%%%
\subsection{Properties of asymmetric MDI-QKD} \label{App:asymmetric:simulation}
We numerically study the properties of an asymmetric MDI-QKD system. In our simulations below, the asymptotic key rate, denoted by $R_{rig}$, is rigorously calculated from the key rate formula given by Eq.~(\ref{Eqn:Key:formula}) in which each term is shown in~\ref{App:QandE}. $R_{est}$ denotes the estimated key rate from Eq.~(\ref{Key:formula:xmuamub}). The practical parameters are listed in Table~\ref{Tab:exp:parameters}. We used the method of~\cite{marcos:finite:2013} for the finite-key analysis.

\begin{figure}[!t]
\centering
\resizebox{7cm}{!}{\includegraphics{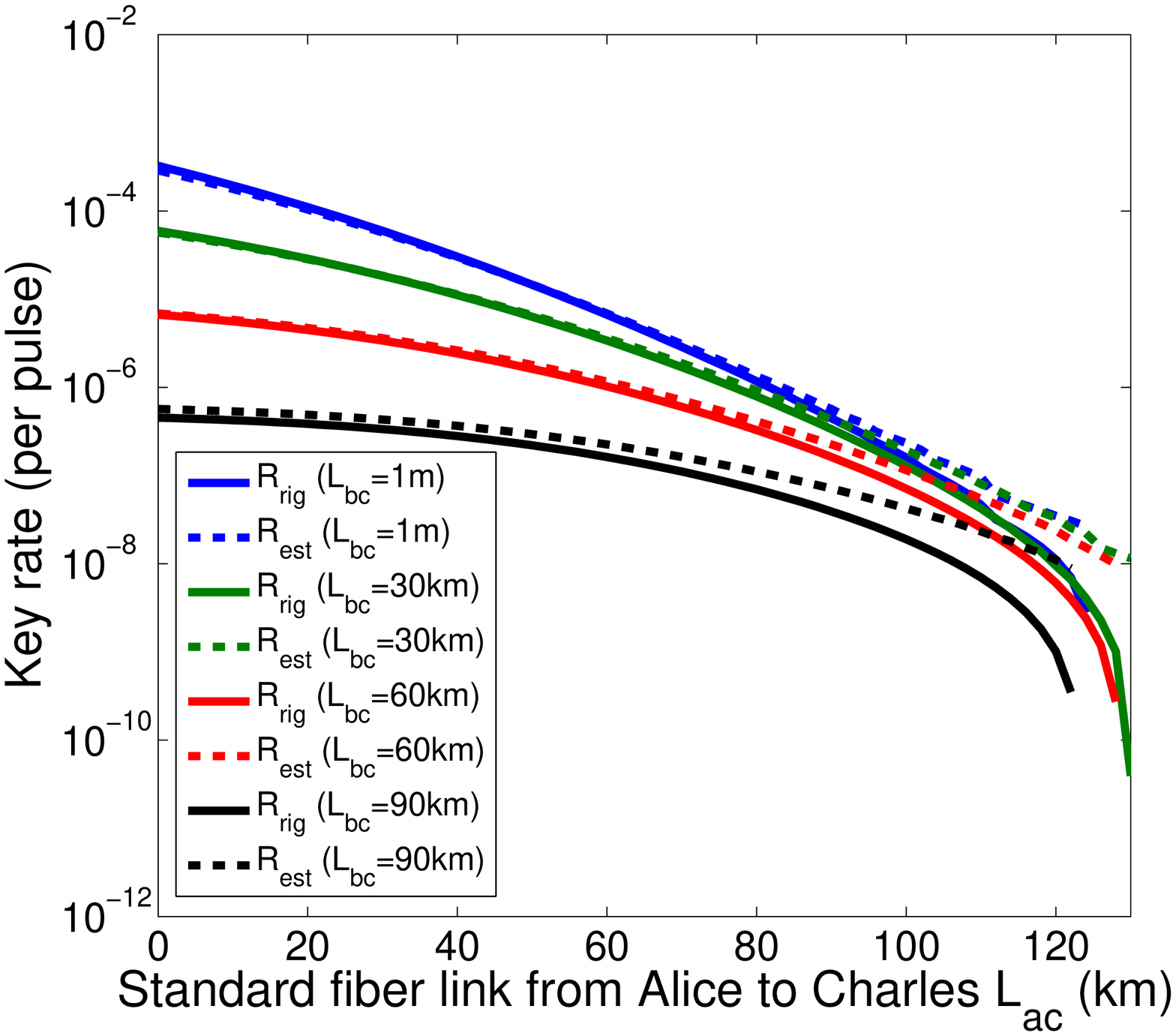}} \caption{\textbf{Asymptotic key rates of $R_{rig}$ and $R_{est}$.} $R_{rig}$ and $R_{est}$ denote respectively the rigorous key rate [Eq.~(\ref{Eqn:Key:formula})] and the estimated key rate (Eq.~(\ref{Key:formula:xmuamub})). At short distances, the overlap between $R_{rig}$ and $R_{est}$ demonstrates the accuracy of our estimation model, while at long distances, background counts affect its accuracy. An asymmetric system can tolerate a maximal channel mismatch of $x$=0.004 (120 km length difference for two standard fiber links). } \label{Simu:Key:Comparision}
\end{figure}
Firstly, Fig.~\ref{Simu:Key:Comparision} simulates the key rates of $R_{rig}$ and $R_{est}$ at different channel lengths. For short distances (\ie, total length$L_{ac}+L_{bc}$$<$100 km), the overlap between $R_{est}$ and $R_{rig}$ demonstrates the accuracy of our estimation model of Eq.~(\ref{Key:formula:xmuamub}). Therefore, in the short distance range, we could focus on $R_{est}$ to understand the behaviors of the key rate. Moreover, from the curve of $L_{bc}$=1m, we have that this asymmetric system can tolerate up to $x$=0.004 (120 km length difference for standard fiber links).

\begin{figure}[!t]
\centering
\resizebox{6.7cm}{!}{\includegraphics{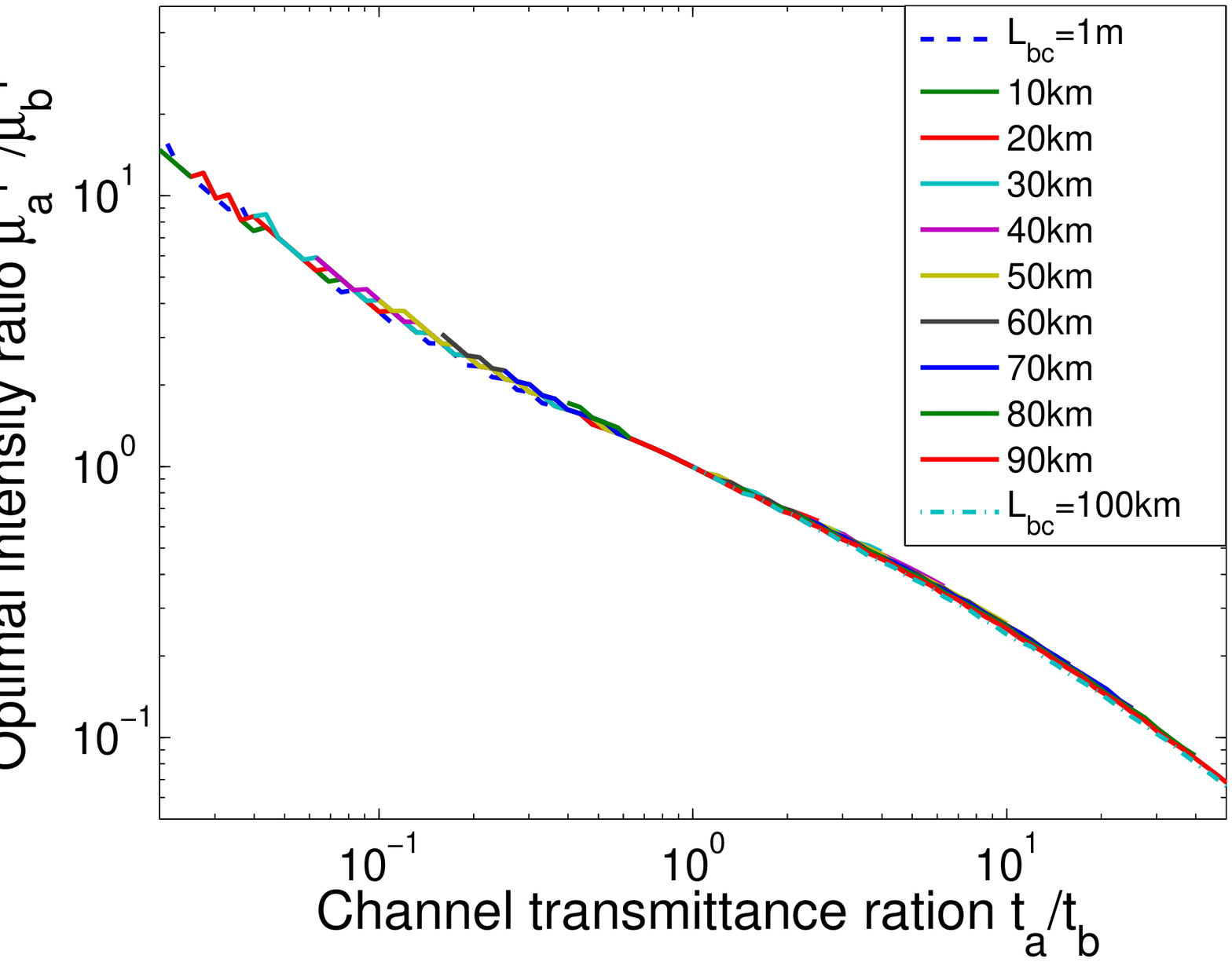}} \caption{\textbf{Optimal $\mu_{a}$ and $\mu_{b}$.} At short distances (\ie, $x$ is around 1 or bigger), $\mu_{a}^{opt}$ and $\mu_{b}^{opt}$ depend \emph{only} on $x$, while at long distances (\ie, $x$$<$0.5), background counts contribute significantly. The non-smooth behaviors here are mainly due to background counts and numerical errors.} \label{Simu:mumu:withdark}
\end{figure}
Secondly, Fig.~\ref{Simu:mumu:withdark} shows $\mu_{a}^{opt}$ and $\mu_{b}^{opt}$, when both $L_{bc}$ and $L_{ac}$ are scanned from 1 m to 100 km. These parameters numerically verify Theorem~\ref{theorem1}: at short distances ($x$$\geq$0.5), $\mu_{a}^{opt}$ and $\mu_{b}^{opt}$ depend \emph{only} on $x$, while at long distances ($x$$<$0.5), background counts contribute significantly and result in non-smooth behaviors. $\mu_{a}^{opt}$ and $\mu_{b}^{opt}$ are both in $O(1)$.

\begin{figure}[!t]
\centering
\resizebox{7cm}{!}{\includegraphics{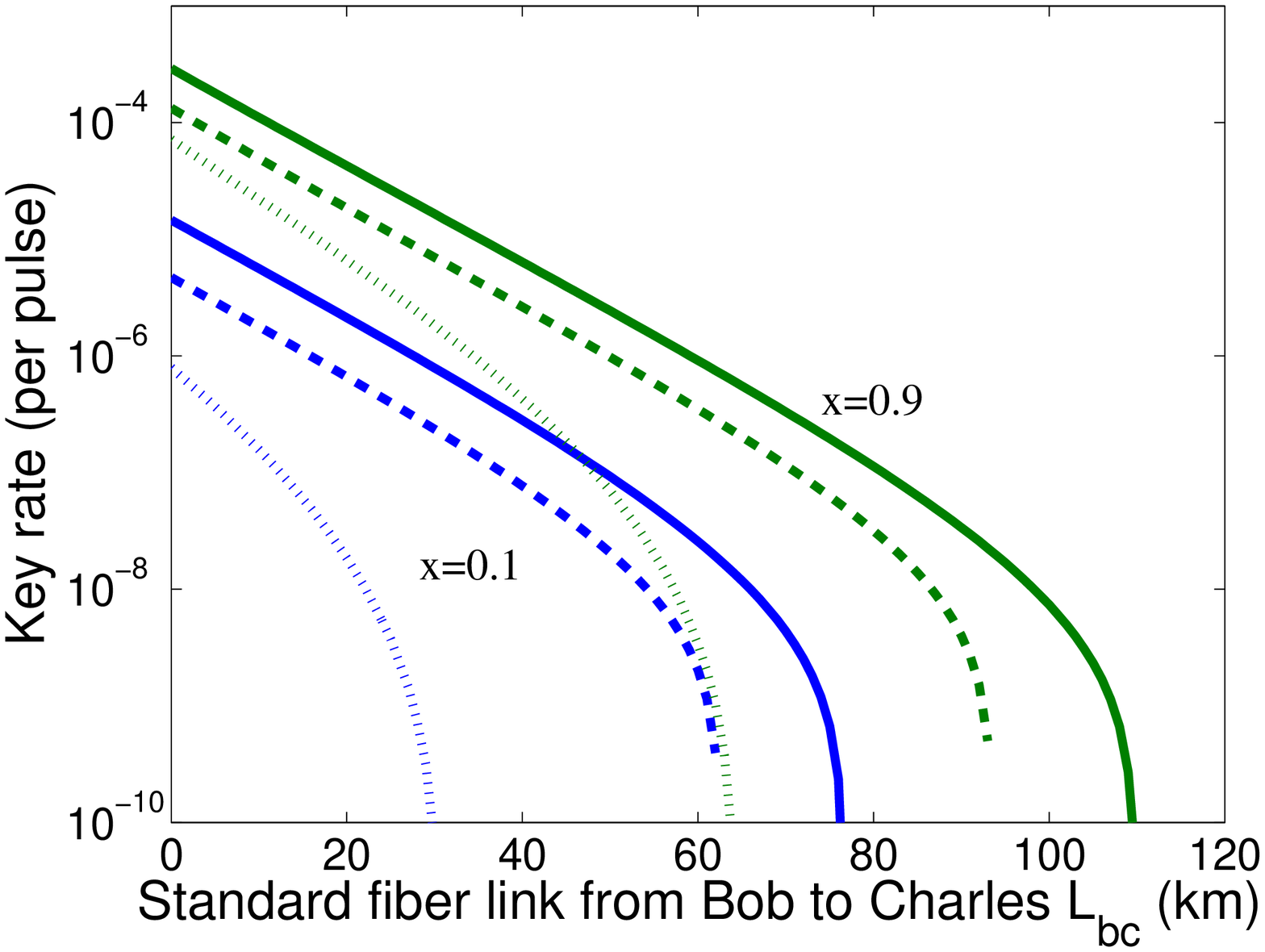}} \caption{\textbf{Key rate with fixed $x$.} Solid curves are the asymptotic key rates: as shown in Eq.~(\ref{Key:formula:xmuamub:log}), $\log_{10}R_{est}$ is linearly proportional to $L_{bc}$. Dashed curves are the two decoy-state key rates without the finite-key effect, \ie, with an infinite number of signals. Dotted curves are the two decoy-state key rates with the finite-key effect: with $x$=0.9, the optimal intensities are $\mu_{a}^{opt}\approx\mu_{b}^{opt}$ and $\nu_{a}^{opt}\approx\nu_{b}^{opt}$; with $x$=0.1, the optimal intensities satisfy $\mu_{a}^{opt}/\mu_{b}^{opt}=\nu_{a}^{opt}/\nu_{b}^{opt}\approx7$ (see Table~\ref{Tab:Opt:Parameters} for some representative values). The weakest decoy state $\omega$ is set to $5\times10^{-4}$.} \label{Fig:key:2decoy:fluc}
\end{figure}
Finally, we simulate the optimal key rates under two \emph{fixed} $x$ in Fig.~\ref{Fig:key:2decoy:fluc}.
\begin{enumerate}
  \item Solid curves are the asymptotic keys: at short distances ($L_{bc}$+$L_{ac}$$<$120 km), the maximal $G(x,\mu_{a},\mu_{b})$ is fixed with a fixed $x$ (see Eq.~(\ref{Key:formula:Gxmuamub})). Taking the logarithm with base 10 of $R_{est}$ and writing $t_{b}$=$10^{-\alpha L_{bc}}$, Eq.~(\ref{Key:formula:xmuamub}) can be expressed as
\begin{equation} \label{Key:formula:xmuamub:log}
    \log_{10}R_{est}=-2\alpha L_{bc}+\log_{10}\frac{\eta_{d}^{2}G(x,\mu_{a},\mu_{b})}{2}. \\
\end{equation}
Hence, the scaling behavior between the logarithm (base 10) of the key rate and the channel distance is linear, which can be seen in the figure. Here, $\alpha=0.2$ dB/km (standard fiber link) results in a slope of -0.4.

  \item Dotted curves are the two decoy-state key rates with the finite-key analysis: we consider a total number of signals $N=10^{14}$ and a security bound of $\epsilon=10^{-10}$; for the dotted curve with $x$=0.1, the optimal intensities satisfy $\mu_{a}/\mu_{b}\approx\nu_{a}/\nu_{b}\approx7$, which means that the ratios for the optimal $\mu$ and $\nu$ are roughly the same and this ratio is mainly determined by $x$. Even taking the finite-key effect into account, the system can still tolerate a total fiber link of 110 km.
\end{enumerate}

\end{appendix}


\begin{thebibliography}{10}
\bibitem{QBERorigin}
From the security aspect, one can in principle operate a MDI-QKD system without knowing the exact origin of the observed QBER. However, from the performance aspect, it is important to study the origin of the QBER in order to maximize the secure key rate.
\bibitem{photonnumber}
We assume that the coherent state is phase-randomized. Thus, its photon number follows a Poisson distribution of mean $\mu$.
\bibitem{QxEx}
In the derivation of $Q_{X}$ and $E_{X}$ with mode mismatch, the non-overlapping modes can be essentially treated as background counts increasing the background count rate of each detector. The final result is a summation over the overlapping and non-overlapping modes.
\bibitem{twodecoy}
For one signal state and two decoy states ($q_a\in\{\mu_a,\nu_a,\omega_a\}$ and $q_b\in\{\mu_b,\nu_b,\omega_b\}$), $Y_{Z}^{1,1}$ and $e_{X}^{1,1}$ can be estimated from the linear equations for 9 intensity pairs.
\bibitem{intensityw}
We assume that, the intensity of the signal state is about 0.5 and the maximum extinction ratio of a practical intensity modulator is around 30 dB~\cite{decoyexp,rosenberg2009practical}. Thus, the lowest intensity that can be modulated is $5\times10^{-4}$.
\bibitem{signaldistribution}
The number of signals in the $X$ (or $Z$) basis and the distribution of the signals over the signal and decoy states are both optimized numerically to maximize the key rate.
\bibitem{differentN}
In more general cases with different numbers of signals \emph{N}, we also find that the optimal choice is around 80\% larger than the symmetric choice.
\end{thebibliography}
\end{document}